\documentclass[manuscript,nonacm,acmsmall]{acmart}
\usepackage{xcolor}
\usepackage{multirow}
\usepackage{array}
\usepackage{graphicx}
\usepackage{acronym}
\usepackage{colortbl}
\usepackage{tikz}
\usepackage{enumitem}
\usepackage{amsthm}
\usepackage{listings}
\usepackage[table]{xcolor}
\usepackage[most]{tcolorbox}

\newtcolorbox{takeawaybox}{
    enhanced,
    breakable,
    colback=gray!5,
    colframe=gray!50,
    boxrule=0.5pt,
    arc=2mm,
    left=8pt,
    right=8pt,
    top=6pt,
    bottom=6pt,
    before skip=10pt,
    after skip=10pt
}

\newcommand*\circled[1]{\tikz[baseline=(char.base)]{
    \node[shape=circle,draw,inner sep=1pt,fill=black] (char) {\textcolor{white}{#1}};}}

\newcolumntype{Q}{>{\setbox0=\hbox\bgroup}c<{\egroup}@{}}
\newcolumntype{L}[1]{>{\raggedright\let\newline\\\arraybackslash\hspace{0pt}}m{#1}}
\newcolumntype{C}[1]{>{\centering\let\newline\\\arraybackslash\hspace{0pt}}m{#1}}
\newcolumntype{R}[1]{>{\raggedleft\let\newline\\\arraybackslash\hspace{0pt}}m{#1}}

\newcolumntype{H}{>{\setbox0=\hbox\bgroup}c<{\egroup}@{}}

\newtheorem{defn}{Definition}
\AtBeginDocument{%
  }

\setcopyright{acmlicensed}
\copyrightyear{2026}
\acmYear{2026}
\acmDOI{XXXXXXX.XXXXXXX}
\acmISBN{978-1-4503-XXXX-X/2018/06}

\begin{document}

\title[Dataset Recommendation via Multimodal Representation Learning]{Dataset Recommendation for Heterogeneous and Dynamic Scholarly Graphs via Multimodal Representation Learning}
\titlenote{This paper has been accepted as a full paper at \emph{ACM Transactions on Recommender Systems}.}
\author{Ornella Irrera}
\orcid{0000-0003-2284-5699}
\email{ornella.irrera@unipd.it}
\affiliation{%
  \institution{University of Padua}
  \streetaddress{Via G. Gradenigo 6/b}
  \city{Padua}
  \country{Italy}
  \postcode{35131}
}
\author{Matteo Lissandrini}
\orcid{0000-0001-7922-5998}
\email{matteo.lissandrini@univr.it}
\affiliation{%
  \institution{University of Verona}
  \streetaddress{via San Francesco, 22 - 37129 Verona}
  \city{Verona}
  \country{Italy}}
  
\author{Daniele Dell'Aglio}
\orcid{0000-0003-4904-2511}
\email{dade@cs.aau.dk}
\affiliation{%
  \institution{Aalborg University}
  \streetaddress{Selma Lagerløfs Vej 300}  
  \postcode{9000}
  \city{Aalborg}
  \country{Denmark}}

\author{Gianmaria Silvello}
\orcid{0000-0003-4970-4554}
\email{gianmaria.silvello@unipd.it}
\affiliation{%
  \institution{University of Padua}
  \streetaddress{Via G. Gradenigo 6/b}
  \city{Padua}
  \country{Italy}
}

\renewcommand{\shortauthors}{Irrera et al.}

\begin{abstract}

Researchers struggle to find relevant datasets amid the volume of data across repositories and the growth of scholarly publications. Manual dataset retrieval is time-consuming and often fails to surface valuable resources not easily discoverable through direct links or popularity metrics. Dataset recommendation — identifying datasets relevant to a paper, author, or query — has become essential for improving dataset discoverability and reuse. Yet this task is difficult because Scholarly Knowledge Graphs (SKGs) are large-scale, heterogeneous, dynamic, and incomplete. Although openly available SKGs organize scientific knowledge by linking scholarly outcomes, they often lack meaningful connections between datasets, leaving them isolated and underutilized. These graphs contain millions of evolving relations but still suffer from missing or inaccurate metadata, duplicate records, and weak dataset integration.

This work presents the Multimodal Scholarly Attention Network (MM-SAN), a heterogeneous multimodal graph representation learning method for recommending datasets relevant to a research publication within real-world SKGs. MM-SAN combines topology-aware and text-based embeddings, enabling effective recommendations even when textual metadata is limited or absent. It operates in transductive, semi-inductive, and fully inductive settings, adapting to evolving SKGs without retraining. Experiments show that MM-SAN surpasses state-of-the-art methods, providing a robust, scalable solution for dataset recommendation.
\end{abstract}



\maketitle

\section{Introduction}
\label{sec:intro}
With the continuous growth of scientific data, finding relevant datasets to validate hypotheses, reproduce experiments, and train algorithms has become increasingly crucial. However, the absence of standardized practices for describing and citing datasets often results in poorly documented resources that are either unlinked or only loosely connected to related research outputs. This lack of descriptive metadata and proper documentation makes datasets difficult to discover and significantly limits their impact within the scientific community~\cite{brickley2019google,Wilkinson2016,silvello_2018,bunemann_2020,bunemann_etal-2021}.  

To improve dataset discoverability, recent methods propose relying on structured and interlinked representations of scholarly information in the form of \acp{SKG}~\cite{lin_etal-2024}. 
\acp{SKG} provide a rich, structured network of scholarly items, such as papers, datasets and authors, and their relationships.
As such, \acp{SKG} form a foundational infrastructure that supports a variety of scholarly tasks, including dataset discovery, information retrieval, and knowledge exploration.

\acp{SKG} like \ac{MAKG}~\cite{farber_2019}, \ac{ORKG}~\cite{jaradeh_etal-2019}, and \ac{OAG}~\cite{manghi_etal-2019}, explicitly include some dataset nodes, yet their information is quite scarce and their interconnections with scientific publications are also extremely limited. 
Datasets are frequently isolated or only loosely linked to related research outputs. 
Additionally, \acp{SKG} are \textit{dynamic}, continuously expanding with new publications and datasets. 
However, they are affected by \textit{incompleteness} (i.e., publications usually contain rich metadata, whereas datasets often lack sufficient descriptions), and by \textit{sparseness}, since links between datasets and publications are relatively weak or rare.
These issues limit the ability of SKGs to represent dataset-related information, thereby making dataset discovery difficult.

In this context, \textit{dataset recommendation}, which involves identifying relevant datasets for a paper, author, or textual query, is crucial for improving dataset discoverability and fostering data reuse. 
However, developing effective dataset recommendation methods requires addressing the inherent challenges of sparse, heterogeneous, dynamic, and incomplete real-world \acp{SKG}.  
Dataset recommendation has to deal with \textit{data sparsity} because, unlike some settings that rely on extensive user-item interaction data, dataset recommendation involves a limited number of data points, making it difficult to identify meaningful patterns~\cite{ChenEtAl2025}. 
Additionally, the \textit{lack of user feedback} eliminates a crucial source of information for refining and validating recommendation models.  
Moreover, dataset recommendation must handle  \textit{heterogeneous and incomplete data}. 
Datasets vary in descriptive metadata quality, and the available information is often inconsistent across different sources, complicating integration and comparison. 
The diversity in data representations and dimensions further exacerbates the challenge of establishing common ground for analysis.  

Finally, dataset recommendation is hindered by a \textit{scarcity of ground-truth labeled data}. 
Verified information for training and evaluation is often limited, restricting large-scale experiments and necessitating reliance on unsupervised or semi-supervised learning approaches. 
While these methods can be effective, they generally do not achieve the accuracy and reliability of supervised learning techniques that depend on abundant labeled data.  

Addressing these challenges requires novel approaches explicitly tailored to dataset recommendation, leveraging the structure of \acp{SKG} while accounting for their limitations.  

State-of-the-art dataset recommendation methods typically address these challenges in isolation, making them poorly suited for real-world applications. 
Most approaches leverage \acp{SKG} to rank relevant datasets~\cite{wang_etal-2022_a,wang_etal-2022_b,altaf_etal-2019,chen_etal-2019} but are evaluated on homogeneous, static, and specialized \acp{SKG}, failing to handle the complexity of real-world graphs. While link prediction on real-world \acp{SKG}~\cite{irrera_etal-2025} may appear as a possible approach, it is inherently limited for recommendation tasks, and no multimodal method currently exists to fully exploit heterogeneous, sparse, dynamic, incomplete SKGs.

As a result, dataset recommendation methods struggle with sparse, dynamic graphs and incomplete descriptions.  
Other methods rely solely on textual metadata from publications and datasets~\cite{viswanathan-etal-2023-datafinder,fabrer_etal-2021}, making them ineffective when metadata are unavailable or of low quality.

In this paper, we tackle the task of dataset recommendation, where the goal is to recommend datasets that are relevant to a given research publication. 
We propose \ac{MM-SAN}, a multimodal heterogeneous graph representation method for dataset recommendation, accommodating heterogeneous, sparse, and incomplete real-world \acp{SKG}. 
\ac{MM-SAN} integrates topology-based and text-based embeddings, exploiting a multimodal learning design, making it effective even when textual features are absent or of low quality.
\ac{MM-SAN} works in transductive and inductive scenarios and is effective whether complete or incomplete textual metadata are available. 
For evaluation, we rely on subsets of \ac{OAG}, the backbone of the European Open Science Cloud (EOSC, \url{https://eosc.eu/}), which is open-access, frequently updated, and contains scientific datasets. 
We address three research questions:
\begin{description}[labelindent=0cm,leftmargin=0cm]
\item[RQ1 (Robustness).] \textcolor{black}{To what extent can MM-SAN provide effective dataset recommendations in the presence of the topological sparsity typical of real-world \acp{SKG}, characterized by limited connectivity between publication and dataset nodes?}
\item[RQ2 (Flexibility).] \textcolor{black}{To what extent can MM-SAN guarantee effective recommendation performance across transductive, semi-inductive, and fully inductive settings, consequently adapting to new, previously unseen nodes without requiring the model to be retrained?}
\item[RQ3 (Versatility).] \textcolor{black}{To what extent can MM-SAN preserve effective recommendation performance under progressively reduced textual metadata availability by leveraging graph topology to compensate for missing semantic information?}

\end{description}
Regarding the first question, we examine how well \ac{MM-SAN} performs in the presence of sparse and incomplete graphs, which are typical in real-world \acp{SKG} scenarios. 
The second research question focuses on \textit{inductiveness}, defined as the model’s ability to adapt to an evolving graph without requiring full retraining. 
In other words, the model is trained on an initial version of the graph and evaluated on an updated version containing new nodes and edges not used for training. 
Hence, we present for the first time a method that is able to generalize across different learning scenarios, i.e., transductive, inductive, and semi-inductive settings.
Lastly, the third question focuses on whether the method can effectively exploit graph structure to mitigate the impact of missing or limited textual metadata.

Our main \textbf{contributions} are the following:
\begin{itemize}[leftmargin=*]
\item A novel multimodal heterogeneous graph representation approach tailored for recommending scientific datasets relevant to research publications, demonstrating strong performance and robustness across transductive, inductive, and semi-inductive scenarios.
\item We provide an extensive evaluation curriculum that for the first time spans across a wide range of settings, from metadata-rich to metadata-free environments. This curriculum allows us to show how \ac{MM-SAN}’s architecture achieves superior performance and sets a new standard for works in this area.
\item We conduct a thorough comparative analysis of \ac{MM-SAN} against state-of-the-art graph-based baselines and a newly introduced competitive text-based baseline, evidencing consistent improvements across all tested scenarios.
\end{itemize}

\textbf{\textit{Outline}.} In Section \ref{sec:related} we describe the baselines; in Section \ref{sec:problem} we formalize the dataset recommendation task; in Section \ref{sec:san} we describe the \ac{MM-SAN} method; in Sections \ref{sec:evaluation} and \ref{sec:results} we describe the experimental setup and discuss the results. In Section \ref{sec:ablation} we report an ablation study. \textcolor{black}{In Section \ref{sec:discussion} we discuss the results and the real-world practical implications.} Finally, in Section \ref{sec:remarks}, we draw some conclusions.

\section{Related work}
\label{sec:related}
In this work, we study methods for dataset recommendation at the intersection of two important areas: \acp{SKG} and graph representation learning.
First, we present \acp{SKG}, which act as the structural backbone for modeling and connecting entities in the scholarly ecosystem, and support tasks like dataset recommendation. 
This overview helps contextualize the structure and characteristics of the data on which dataset recommendation methods operate. 
Second, we examine Graph Representation Learning (GRL) methods, which enable the extraction of informative and expressive node embeddings from \acp{SKG}. 
These methods play a crucial role in various downstream tasks, including link prediction, node classification, and recommendation systems, among others. 
Finally, we review existing approaches that have been applied to the dataset recommendation task.

\paragraph{\textbf{Scholarly knowledge graphs.}} 
\acp{SKG} organize and structure scholarly knowledge, providing a means to access, discover and explore outcomes such as publications, datasets, software, authors and venues. 
Several \acp{SKG}, like AMiner~\cite{tang2008arnetminer} and DBLP~\cite{ackermann_etal-2024}, still focus mainly on representing connections among publications and related authors and do not include dataset nodes.


Conversely, an increasing number of \acp{SKG} explicitly model datasets as first-class entities. The most recent example is the \ac{MES} graph~\cite{irrera_etal-2023} which has been released as a dataset-centric subgraph of \ac{OAG}; thanks to its semi-automatic curation process, it offers a high-quality, reliable ground-truth resource that is particularly suited for evaluating methods on dataset discovery and related tasks.
Other prominent examples include \ac{MAKG}~\cite{farber_2019}, which comprises roughly 200M nodes and more than 10B edges; \ac{ORKG}, with about 5M nodes and 50M relationships; \ac{DSKG}~\cite{farber2021data}, which links around $2K$ datasets to $635K$ publications and connects them to other Linked Data sources such as \ac{MAKG}, ORCID, and Wikidata; and \ac{OAG}~\cite{manghi_etal-2019}, the largest of these graphs, with approximately 227M nodes and 15B relationships, openly available and including over 60M datasets.

Existing evaluations of dataset recommendation have largely relied on narrow, ad-hoc collections tailored to metadata-based prediction and recommendation, which in many cases do not constitute full-fledged \acp{SKG}. Consequently, in this work we deliberately exclude the following resources:
\begin{itemize}
    \item the LinearSVM\_Dataset, a bipartite graph built from $1{,}691$ dataset titles derived from \ac{DSKG}~\cite{farber2021data} and 88K publication abstracts from \ac{MAKG}, which is specifically engineered for a particular experimental setup;
    \item the DataFinder dataset~\cite{viswanathan-etal-2023-datafinder}, which cannot be regarded as a proper \ac{SKG} because it omits key node types such as authors, venues, and organizations, and instead only provides 17K textual queries paired with 7K relevant datasets;
    \item the Delve~\cite{akujuobi_etal-2017delve} subsets used to evaluate the HVGAE~\cite{altaf_etal-2019} and AMENDER~\cite{chen_etal-2019} baselines, whose graph structure is unusually dense and therefore unrepresentative of the typically sparse nature of real-world \acp{SKG} where datasets are usually connected to at most five publications \cite{irrera_etal-2024}.
\end{itemize}

\paragraph{\textbf{Graph representation learning methods.}}
\ac{GRL} provides a general paradigm in which graph-structured data are mapped into low-dimensional continuous vector spaces, enabling the learning of expressive node embeddings that capture both structural and semantic properties of the graph. These representations are then exploited by a wide range of methods to support downstream tasks such as node classification, link prediction, and recommendation.

\acp{GCN}\cite{kipf_etal-2020} represent one of the foundational architectures for learning node embeddings, where each layer updates a node’s representation by propagating and combining information from its neighbors through graph convolutions. For homogeneous graphs, GraphSAGE \cite{hamilton_etal-2017} generalizes this idea by sampling a fixed number of neighbors per node and aggregating their features via different schemes, such as mean, LSTM-based, or pooling aggregators. GAT~\cite{velivckovic_etal-2017} advances this paradigm by using attention mechanisms to assign adaptive weights to neighboring nodes, leading to more expressive embeddings and strong performance, particularly in inductive learning scenarios.

When dealing with heterogeneous graphs that contain multiple node types, \ac{GRL} methods must cope with additional complexity. HetGNN~\cite{zhang_etal-2019} tackles this by sampling type-specific neighbors from each node’s neighborhood and combining their information through a bi-LSTM-based aggregator. RGCN~\cite{chen2019rgcn} generalizes \acp{GCN} by explicitly modeling edge and node types, thereby enabling more expressive relational representations, \textcolor{black}{while KGCN~\cite{wang_etal-2019kgcn} aggregates neighborhood information in a knowledge graph using biased, user-specific relation weights to capture both structural and semantic relatedness.}

Other techniques leverage attention mechanisms to handle heterogeneous graphs better. GATNE~\cite{cen_etal-2019} tackles multiplex graphs by integrating edge attributes, while HGT~\cite{hu_etal-2020} uses a transformer-based, type-specific attention mechanism to manage diverse node and edge types. \textcolor{black}{KGAT~\cite{wang_etal-2019kgat} applies attentive propagation over a unified collaborative knowledge graph to weigh neighbor contributions by relation-aware importance.} HiNormer~\cite{mao2023hinormer} introduces two encoders -- a local structure encoder and a heterogeneous relation encoder -- tailored to handle the complexity of heterogeneous graphs.
Methods like HAN~\cite{wang_etal-2022han} and MAGNN~\cite{fu_etal-2020} utilize \emph{metapaths}, which are sequences of node and edge types that define relational patterns. HAN aggregates information from different metapaths, whereas MAGNN performs intra- and inter-metapath aggregation.
These baselines have been tested on various heavily cleaned AMiner and DBLP subsets without considering datasets. SAN~\cite{irrera_etal-2025} is another \ac{GRL} approach tailored to link prediction in real-world \acp{SKG}, which performs random-walk-based neighbor sampling and then aggregates the sampled information using an attention mechanism.

\paragraph{\textbf{Dataset recommendation methods.}}
In recent years, dataset recommendation has garnered significant attention, leading to specialized methods for this task. Most of the proposed approaches focus on recommending scientific datasets that are relevant to research publications.
Some approaches focus on leveraging textual metadata from publications and datasets to generate recommendations. For instance, LinearSVM~\cite{fabrer_etal-2021} recommends datasets based on a lengthy textual description, such as the abstract of a paper. Another approach introduced a neural bi-encoder model~\cite{viswanathan-etal-2023-datafinder} to recommend datasets given a set of publication abstracts.
Other approaches focus on graph exploration to identify relevant datasets. 
~\citet{wang_etal-2022_a,wang_etal-2022_b}
combined citation and co-author network exploration in \ac{MAKG} to recommend both publications and datasets. 
However, this approach considered a dataset with fewer than 1K datasets, thus the produced output seldom included datasets.
HVGAE~\cite{altaf_etal-2019} instead adopts a heterogeneous graph variational autoencoder. 
Similarly, AMENDER~\cite{chen_etal-2019} introduces an attentive multi-layered network for recommending datasets to authors. HVGAE and AMENDER have been tested on two subsets of the Delve dataset~\cite{akujuobi_etal-2017delve}, extracted and processed ad-hoc for these tasks. HetGNN~\cite{zhang_etal-2019} has been used for author-to-venue recommendations, which are more straightforward than dataset recommendations because venues are well-connected in a \ac{SKG}, providing valuable information that aids recommendations. There are fewer venues than publications and datasets, simplifying the final ranking task.
Most dataset recommendation methods have been evaluated only on dense, ad-hoc graphs, which limits their applicability to more complex, real-world \acp{SKG}. This is especially problematic since comprehensive textual descriptions of datasets are often unavailable. In \cite{irrera_etal-2024}, we reproduced these methods and found them less effective in real-world scenarios. Our analysis also tested the generalizability of the dataset recommendation task using three popular recommendation baselines (TopPop \cite{cremonesi_etal-2012}, BPR \cite{rendle_etal-2009}, and LightGCN \cite{he_etal-2020}), underscoring the need for a new, robust, and versatile method that can perform well in real-world settings.

\textcolor{black}{MM-SAN, introduced in this work, is designed to overcome the limitations of current dataset recommendation methods. Unlike existing approaches, MM-SAN is evaluated on real-world scholarly datasets, a setting that has so far been largely overlooked in this domain. Moreover, its ability to operate effectively in inductive scenarios distinguishes it within the broader landscape of graph representation learning.
Previous work on real-world \acp{SKG} has shown that graph-based models can be effective for the related task of link prediction \cite{irrera_etal-2025}. Motivated by this evidence, MM-SAN tackles the problem of dataset recommendation by leveraging a multimodal setting tailored to this task.}




\section{The dataset recommendation task}
\label{sec:problem}
In what follows, we introduce the formal definitions of \acp{SKG} and heterogeneous \ac{GRL}, and we characterize the dataset recommendation task together with the main open challenges it poses.
\begin{defn}
A \acf{SKG} is a tuple
\[ \mathcal{G}{:}(\mathcal{V}, \mathcal{E},\mathcal{A}, \mathcal{R}, \mathcal{T}_\mathcal{V}, \alpha, \beta, \phi_\mathcal{V}) \] 
where \( \mathcal{V}, \mathcal{E},\mathcal{A}, \mathcal{R}, \mathcal{T}_\mathcal{V} \) 
are the sets of nodes, edges, node types, edge types, and node attributes respectively. 
Then, \( \alpha{:}\mathcal{V}{\to}\mathcal{A} \)  is a node type function that maps each node \( v{\in}\mathcal{V}\) to one node type \( a{\in}\mathcal{A} \),  \( \beta{:}\mathcal{E}{\to}\mathcal{R} \) is an edge type function that maps each edge \( e{\in}\mathcal{E}\) to an edge type \( r{\in}\mathcal{R} \), and \( \phi_\mathcal{V}{:}\mathcal{V}{\to}\mathcal{\mathcal{T}_\mathcal{V}} \) is a function that maps each node to a textual attribute  \( t{\in}\mathcal{\mathcal{T}_\mathcal{V}}\). 
A \ac{SKG} is  \emph{heterogeneous} if it has multiple node and edge types such that \( \mathcal{|A|} + \mathcal{|R|} > 2 \). 
\end{defn}

Generally, an \ac{SKG} incorporates both structural (edges) and semantic (nodes, their types, and their attributes) information, where the attributes are the textual metadata of a scholarly entity. This provides a rich representation for the dataset recommendation task. Its heterogeneous nature adds complexity and allows for modeling the graph's diverse relationships and attributes between various entities.

\begin{defn}
\textcolor{black}{Given a heterogeneous \ac{SKG} \( \mathcal{G} \), where $\mathcal{V} = \mathcal{V}_p \cup \mathcal{V}_d$ represents the set of publication nodes $\mathcal{V}_p$ and dataset nodes $\mathcal{V}_d$, and \( q {\in} \mathcal{V}_\text{p} \) is  the \emph{query node}, the \textbf{dataset recommendation task} aims  to learn a scoring function \( \rho: \mathcal{V}_\text{p} \times \mathcal{V}_\text{d} \to \mathbb{R} \). This function assigns a relevance score for each dataset node \( d_k{\in}\mathcal{V}_\text{d} \) to the query node \( q \). 
The task output is a ranked list \( \pi_q = (d_{(1)}, d_{(2)}, \dots, d_{(n)}) \), with $n = |\mathcal{V}_\text{d}|$, obtained by ordering \( \mathcal{V}_\text{d} = \{d_1, \dots, d_n\} \) according to \( \rho(q, \cdot) \), such that
\[
\rho(q, d_{(1)}) \geq \rho(q, d_{(2)}) \geq \dots \geq \rho(q, d_{(n)}).
\]}

\end{defn}

In this work, the dataset recommendation task is defined as identifying and ranking datasets that are most relevant to a specific research publication, which serves as the query.

\begin{defn}
Given a heterogeneous \ac{SKG} \( \mathcal{G} \), the objective of \ac{GRL} is to learn a function \( f{:}\mathcal{V}{\to}\mathbb{R}^d \) that maps each node \( v{\in}\mathcal{V} \) to a \( d \)-dimensional embedding space, where \( d{\ll}|\mathcal{V}|\). This embedding is designed such that for any two nodes \( u \in \mathcal{V}_\text{p} \) and \( v \in \mathcal{V}_\text{d} \), the dot-product of their corresponding vectors $\vec{u} \cdot \vec{v}$ approximates the scoring function \( \rho \).
\end{defn}

\ac{GRL} provides a way to map heterogeneous nodes into a shared embedding space, capturing their features and the graph's structure. This is crucial for the dataset recommendation task, where the relationship between query nodes (publications) and potential datasets must be effectively learned and represented in the embedding space.


\paragraph{\textbf{\textcolor{black}{Challenges.}}} The dataset recommendation task faces significant challenges due to the nature of \acp{SKG}. These graphs often contain multiple sparse connected components, duplicated nodes, missing links, and heterogeneous incomplete metadata. Moreover, the extreme class imbalance in ranking exacerbates the task's complexity. In real-world \acp{SKG}, the number of datasets can reach millions, but only a small subset is relevant to a given publication \cite{irrera_etal-2024}. As a result, the task involves accurately retrieving a small set of relevant datasets from a large pool of irrelevant ones.

\textcolor{black}{The first key challenge concerns the \textit{heterogeneity} and \textit{incompleteness} of the metadata available in \acp{SKG}. Publications and datasets are typically deposited across different repositories, each adopting its own metadata schema, controlled vocabulary, and level of descriptive detail. As a result, even when a textual attribute is present for a node, its content can vary substantially in structure, length, and informativeness — ranging from a rich abstract to a bare title, or to a short, non-descriptive label such as a repository-generated identifier. This inconsistency is particularly pronounced for datasets: while publications are subject to relatively standardized editorial and peer-review processes that ensure a minimum level of descriptive quality, datasets are often deposited with minimal or inconsistent documentation. This heterogeneity also complicates the integration of information across sources, as textual descriptions referring to conceptually similar resources may differ substantially in vocabulary, granularity, and completeness, making it difficult to establish a common ground for comparison. Addressing this challenge requires methods that do not depend exclusively on the availability or quality of textual metadata, but can instead complement it with alternative, structurally-derived sources of evidence.}

\textcolor{black}{Another challenge in dataset recommendation is the \textit{lack of user feedback}. In conventional recommender systems, relevance signals typically arise from direct user–item interactions, such as clicks, purchases, ratings, or reviews. These interactions are continuously generated and provide abundant, relatively inexpensive supervision for both model training and evaluation. In contrast, publication–dataset relationships in scholarly graphs emerge through a much slower and less systematic process. A dataset is first produced and deposited, usually with associated metadata, and may subsequently be used, mentioned, or formally referenced by a publication. However, the presence of such a reference in a publication does not necessarily result in a publication–dataset link being recorded in the \ac{SKG}. In some cases, publication–dataset relationships are deposited by researchers; in most cases, however, they are derived through automated extraction procedures applied to the textual content of scientific publications. Whether a dataset mention is successfully identified and converted into a formal graph edge depends on how the dataset is referred to in the text and on the structure and recognizability of the reference. Consequently, a publication may  use or mention a relevant dataset while the corresponding relationship remains unrecorded in the graph. The observed graph should therefore be regarded as an incomplete representation of the underlying publication–dataset relationships, potentially containing only a fraction of the links that exist in reality, thereby leading to a high \textit{sparsity} of publication and dataset nodes, and to the \textit{scarcity of ground-truth labeled data for evaluation}.}

\textcolor{black}{These aspects affect both learning and evaluation. Since each publication is associated with only a small number of observed datasets, the amount of positive supervision available for learning robust publication–dataset associations is inherently limited. At the same time, the incomplete and sparse nature of the graph introduces a substantial false-negative problem: a dataset predicted by the model may in fact be relevant to a publication, while being treated as incorrect simply because the corresponding relationship has not been extracted or recorded.}

\textcolor{black}{This phenomenon is related to, but importantly distinct from, the long-tail problem commonly discussed in recommender systems. In conventional long-tail settings, the majority of interactions are concentrated around a relatively popular and densely connected head, while recommendation becomes progressively more difficult for less frequently interacted users or items. The head nevertheless provides substantial training and evaluation signals that models can exploit. Our setting differs fundamentally in that sparsity is pervasive throughout the \ac{SKG}. Both publications and datasets are associated, on average, with at most one counterpart on average in the opposite set. Thus, the challenge is not restricted to a relatively small population of tail nodes; rather, the entire publication–dataset graph is characterized by extremely limited observed connectivity. There is consequently no reliably observed head from which models can derive abundant interaction patterns, and the incompleteness of the graph may affect both training and evaluation. }

\section{MM-SAN}
\label{sec:san}
\begin{figure*}[t]
\includegraphics[width=1\textwidth]{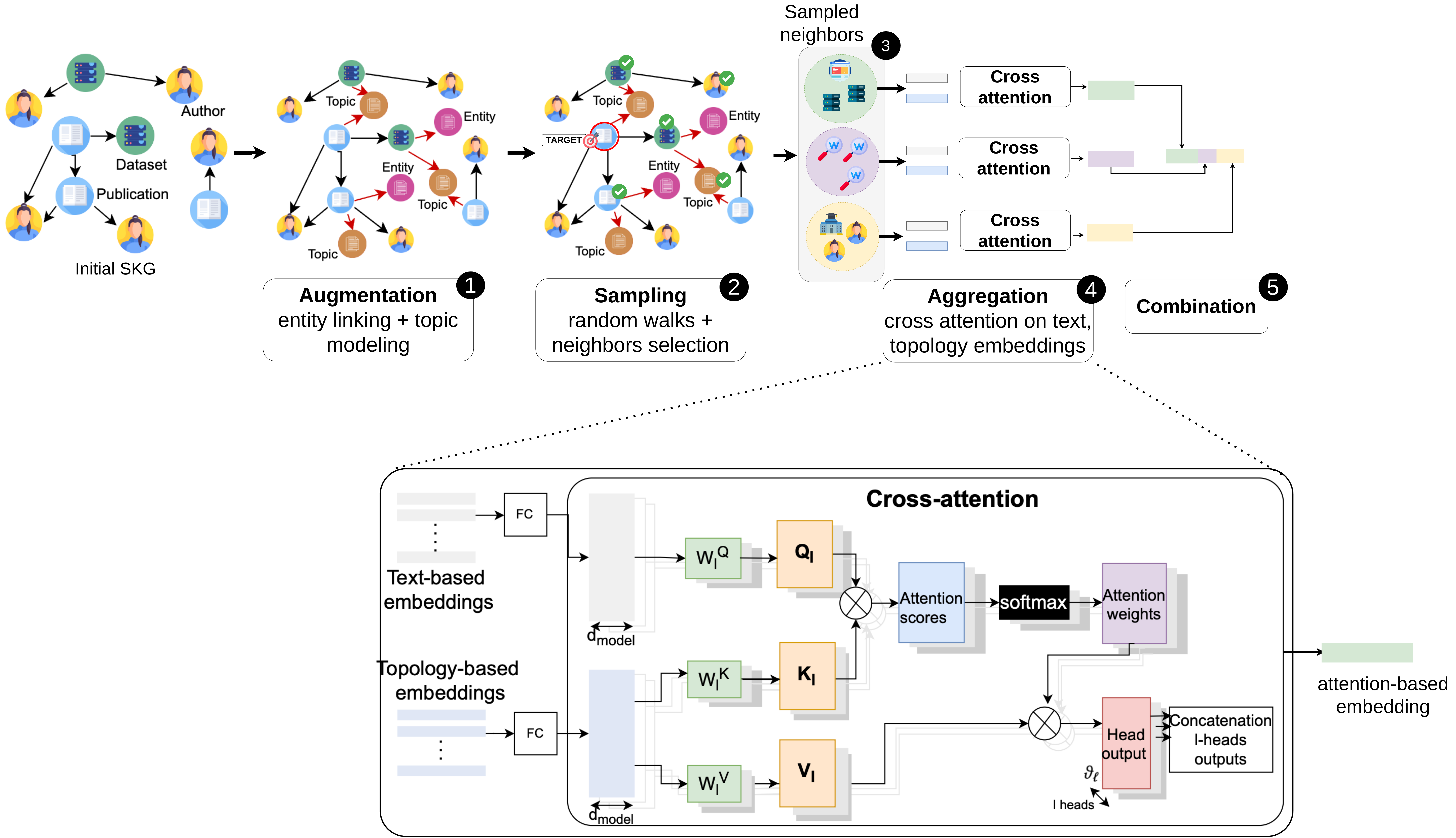}
\caption{The MM-SAN Architecture comprising the augmentation, neighbors sampling, and the aggregation phases. 
\textcolor{black}{\textcircled{1}~\emph{Augmentation}: the input SKG is enriched offline with \emph{topic} and \emph{entity} nodes derived from entity linking and topic modeling, to mitigate structural sparsity. 
\textcircled{2}~\emph{Sampling}: starting from a target node, a set of random walks is generated to explore its local neighborhood. 
\textcircled{3}~\emph{Sampled neighbors}: from the collected walks, MM-SAN selects a representative and type-balanced set of neighbors, grouped into three categories according to their similarity in metadata -- (i) publications and datasets, (ii) keywords, topics, and entities, and (iii) authors, venues, and organizations. 
\textcircled{4}~\emph{Aggregation}: for each of the three groups, a cross-attention module combines the text-based and topology-based embeddings of the sampled neighbors to produce group-specific attention-based embeddings (see the cross-attention detail at the bottom of the figure). 
\textcircled{5}~\emph{Combination}: the three group embeddings are concatenated to form the final representation of the target node.}}\label{fig:san_architecture}
\Description{The MM-SAN Architecture.}
\end{figure*}
This section presents the \ac{MM-SAN} architecture illustrated in Figure \ref{fig:san_architecture} comprising the augmentation, neighbor sampling, and aggregation phases. 
\textcolor{black}{Prior work has explored a graph-based model for the link prediction task \cite{irrera_etal-2025}, where the underlying augmentation, sampling, and aggregation pipeline proved effective on real-world scholarly graphs. Motivated by these results, in this work we target the dataset recommendation task: we adopt a similar augmentation and sampling strategy, and introduce a novel multimodal aggregation mechanism that explicitly accounts for the contribution of graph topology and textual information, while remaining robust to the absence of either modality.}

\paragraph{\textbf{\textit{Augmentation}}}
The augmentation phase (\circled{1}) is executed offline to alleviate the sparsity that typically characterizes \acp{SKG}, thereby improving \textit{robustness (RQ1)} by injecting additional nodes and edges that densify the graph.
Starting from an input \ac{SKG}, \ac{MM-SAN} first collects the textual metadata available for \textit{publication} and \textit{dataset} nodes. It then applies entity linking~\cite{balog_2018}, i.e., the process of detecting mentions of real-world entities in text and aligning them with entries in a knowledge base, which helps disambiguate  entity mentions. In parallel, topic modeling is used to collect the main themes present in the corpus of textual descriptions. For this, we rely on BERTopic~\cite{grootendorst_2022}, which combines transformer-based embeddings with c-TF-IDF to cluster documents according to their shared topical structure. \textcolor{black}{We selected \text{BERTopic} for the topic modeling step because it provides an effective combination of semantic quality, ease of use, and computational efficiency on heterogeneous corpora of varying length. These properties are particularly relevant in our setting, where the augmentation phase must remain scalable to large graphs such as PubMed.}

Within this framework, topics capture high-level themes that act as hubs in the \ac{SKG}, connecting fragments of the graph that would otherwise remain weakly or not at all related. Entities, instead, are fine-grained signals: when multiple nodes mention the same entity, this suggests that they focus on closely related or overlapping content. Exploiting this complementarity, topics are primarily used to promote global connectivity, while entities emphasize local, content-specific relationships between nodes. Both are then injected into the \ac{SKG} as new nodes of type \textit{topic} and \textit{entity}, respectively.

After enrichment, \ac{MM-SAN} derives for each node two distinct representations: a structural embedding learned from the graph topology via \texttt{node2vec}~\cite{grover_etal-2016}, and a textual embedding computed from its metadata. For the latter, we employ the \texttt{all-MiniLM-L6-v2} sentence transformer\footnote{\url{https://huggingface.co/sentence-transformers/all-MiniLM-L6-v2}} for longer textual descriptions and \texttt{Phrase-BERT}~\cite{wang-etal-2021-phrase} for shorter texts such as topic labels.

\textcolor{black}{For structural information, we adopt \texttt{node2vec}, a well-established and efficient approach for learning topological representations from large graphs. Its random-walk parameters ($p$, $q$) provide control over the balance between local and more distant neighborhood exploration, which is  useful for our \acp{SKG}, characterized by tens of thousands of nodes and highly imbalanced node types.}

\textcolor{black}{For textual information, we distinguish between long and short textual attributes. Longer descriptions, including publication and dataset titles and abstracts, are encoded using \texttt{all-MiniLM-L6-v2}, a lightweight sentence-transformer that provides a favorable trade-off between semantic representation quality and computational cost. Its widespread adoption for semantic similarity tasks also facilitates reproducibility and comparison with existing approaches. 
In contrast, short textual attributes, such as topics, entities, and keywords, typically consist of only one to five tokens. These are therefore encoded using \texttt{Phrase-BERT}, which is specifically designed to capture the semantics of short phrases and is better suited to this setting than general-purpose sentence encoders optimized for longer textual inputs.}
We validated these choices by replacing \texttt{node2vec} with \texttt{DeepWalk}~\cite{perozzi_etal-2014} and the transformer-based text encoders with mean-pooled GloVe embeddings~\cite{glove} (Appendix~\ref{app:encoder}, Table~\ref{tab:simpler}). The simpler encoders consistently reduce effectiveness, with the performance gap widening in the inductive setting with partial metadata. This indicates that richer encoders play a key role when limited information is available about the target nodes, enabling them to be represented through more informative and representative embeddings.

\paragraph{\textbf{Sampling.}}
Starting from a target node $v_t$ in the enriched \ac{SKG}, phase \circled{2} focuses on exploring its local neighborhood and selecting a representative set of neighbors across the different node types, while keeping their counts as balanced as possible.
A straightforward strategy is to sample neighbors uniformly at random, as done in methods such as GraphSAGE. Yet, in heterogeneous \acp{SKG} this often leads to poorly representative samples, since node-type distributions are typically highly skewed. For example, author nodes are frequently under-disambiguated, giving rise to many weakly connected or isolated nodes, whereas venue nodes are fewer but usually more densely connected. As a result, approaches like random sampling tend to over-sample authors and under-sample other crucial node types, like venues, producing a biased and less diverse neighborhood.
To mitigate this problem, we adopt a neighbor selection scheme based on random walks.

For a given target node  \( v_t \), we generate a set of random walks of length \( l \).
We ignore edge directions because for each edge, we assume that there exists an edge in the opposite direction (e.g., for each \texttt{cites} edge, we can assume a \texttt{cited by} exists in the opposite direction).
We collect the \( k \) random walks with the highest similarity scores with the target node \( v_t \), computed as follows:
\[
\text{sim\_score}(v_t,\text{walk}_j) = \frac{1}{m_j}\sum_{i{\in}\text{walk}_j} \frac{\textit{cos}(\mathbf{v}_{t}, \mathbf{v}_{i})}{d(v_t,v_i)}
\]
where \( \text{walk}_j \) is one random walk, \( m_j \) is the number of publication and dataset nodes in the walk \( j \), 
 \( v_i \in \{\mathcal{V}_{\text{p}} \cup \mathcal{V}_{\text{d}}\} \) is either a \textit{publication} or a \textit{dataset} node,
\( \mathbf{v}_{t} \) and \( \mathbf{v}_{i} \) are the embeddings encoding the textual metadata of \( v_t \) and \( v_i \) respectively, 
\( \textit{cos}(\mathbf{v}_{\text{t}}, \mathbf{v}_{i}) \) is the cosine similarity defined as \( \frac{\mathbf{v}_{\text{t}} \cdot \mathbf{v}_{i}}{\|\mathbf{v}_{\text{t}}\| \|\mathbf{v}_{i}\|}\), and \( d \) is the distance measured as the number of hops between the target \( v_t \) and \( v_i \) in the walk. 
This formula ensures that publications and datasets that are both similar and closer are given higher priority than those that are similar but more distant.

In step \circled{3}, \ac{MM-SAN} selects the top \( n \) publications and datasets with the highest cosine similarity to the target \( v_t \) from the \( k \) selected walks. For all other node types, it selects the \( n \) nodes that are closest to \( v_t\) within each type. 
The central concept is that entities and topics are essential for revealing previously disconnected parts of the graph, enabling a more thorough exploration of the target node's neighborhood. In contrast, publication and dataset nodes are important for deciding which walks to prioritize as they provide insights into content similarity. 

The resulting set of neighbors can be divided into three groups: (i) one containing publications and datasets, (ii) one containing keywords, topics, and entities, and (iii) one containing authors, venues, and organizations. This classification is based on the similarity of the metadata of the nodes within the same group: publications and datasets share titles and descriptions; keywords, topics, and entities are sets of words that describe a research outcome; and proper nouns often identify authors, venues, and organizations.
In the sampling phase, we aim to improve \textit{flexibility (RQ2)} by focusing on a subset of representative neighbors, rather than learning a global representation for the entire neighborhood of a node. This approach enables the model to adapt to dynamic graphs of different sizes without needing global information about all the nodes.

\paragraph{\textbf{Aggregation.}}
The aggregation step aims to construct the final embedding representation of the target node \( v_t \) by aggregating the neighbor vectors. In particular, this step is essential to ensure that the absence or incompleteness of metadata for \( v_t \) and its selected neighbors does not affect its final embedding. In this phase, \ac{MM-SAN} integrates heterogeneous multimodal data to generate robust node embeddings. This step takes inspiration from classical multimodal settings, where different data types, such as text and images, are considered together. Similarly, MM-SAN treats textual and graph topological information as two distinct modalities, each contributing to generating the final node representation.

The aggregation process (\circled{4}) ensures that even when metadata is missing or incomplete for a node  
$v_t$ or its neighbors, the model can still generate a meaningful embedding for $v_t$. This is achieved by utilizing cross-attention~\cite{vaswani2017attention} to effectively combine neighbor embeddings based on both textual and topological features. Cross-attention allows the model to integrate information from both modalities; for instance, if a node has sparse connections in the graph, the model can rely more on textual embeddings to enrich the representation. Conversely, if textual metadata is absent, topological information from the graph structure plays a stronger role. This approach mirrors the flexibility found in traditional multimodal settings, where the model integrates diverse data sources to make informed predictions. Hence, this phase enhances \textit{versatility (RQ3)} by tackling the integration of diverse information sources, allowing the model to adapt to various real metadata scenarios.

MM-SAN first linearly projects the text-based and topology-based embeddings (FC blocks in Figure \ref{fig:san_architecture}). Then, \ac{MM-SAN} combines three input matrices: Query ($\mathbf{Q}$), Key ($\mathbf{K}$), and Value ($\mathbf{V}$), of sizes  \((n_Q, d_{model})\), \((n_K, d_{model})\), \((n_V, d_{model})\) where $n_Q$, $n_K$, $n_V$ are the numbers of queries, keys, values embeddings respectively, and $d_{model}$ their dimension. 
The queries are the linear projected text-based embeddings while keys and values are 
the linearly projected topology-based embeddings of the selected neighbor nodes. 
For each attention head
${\vartheta}_0,\dots,{\vartheta}_{\ell},\dots,{\vartheta}_H$, 
these matrices are transformed as follows: $\mathbf{Q}_{\ell}{=}\mathbf{Q} \mathbf{W}_{\ell}^Q$, $\mathbf{K}_{\ell}{=}\mathbf{K} \mathbf{W}_{\ell}^K$, and $\mathbf{V}_{\ell}{=}\mathbf{V} \mathbf{W}_{\ell}^V$, where $\mathbf{W}_{\ell}^Q$, $\mathbf{W}_{\ell}^K$, and $\mathbf{W}_{\ell}^V$ are projection matrices of size $(d_{model}, d_k)$ and $d_k{=}\frac{d_{model}}{H}$ is the dimensionality of the projection space for each head. The embeddings for \( \mathbf{Q} \) are based on text, while the embeddings for \( \mathbf{K} \) and \( \mathbf{V} \) are based on topology.
For each attention head ${\vartheta}_{\ell}$ the attention scores are computed using the scaled dot-product attention mechanism, and: 
\[{\vartheta}_{\ell} = \text{softmax}\left( \frac{\mathbf{Q}_{\ell} \mathbf{K}_{\ell}^T}{\sqrt{d_k}} \right) \mathbf{V}_{\ell}.\] 
The outputs from all heads are then concatenated to form the final output: 
\[\text{CrossAttention}(\mathbf{Q}, \mathbf{K}, \mathbf{V}) = \text{Concat}({\vartheta}_0,\dots,{\vartheta}_{\ell},\dots,{\vartheta}_H).  \] 
The result is a matrix of size $(n_Q, d_{\text{model}})$ combining information from different representation spaces.
Cross-attention is applied independently to the three groups of node types: (i) publications and datasets, (ii) keywords, entities, and topics, and (iii) authors, venues, and organizations.
The embeddings resulting for each group are finally concatenated (\circled{5}) to obtain the final target node embedding representation. 

We experimented with various embedding combination methods, including multihead attention~\cite{vaswani2017attention}, bi-LSTM~\cite{graves_etal-2008}, and GRU~\cite{chung_etal-2014}, both as replacements for cross-attention and for the final concatenation step. These results are presented in the ablation study (cf. Sec. \ref{sec:ablation}). However, cross-attention excels at capturing the intricate relationships between topological structures and textual metadata, allowing the model to understand how individual node features interact across different nodes and their descriptions. This capability is especially crucial in heterogeneous graphs, where different types of nodes may have varying levels of importance.

\paragraph{\textbf{Training.}}
To train the model, we employed a mini-batch gradient descent approach. The optimization was done by minimizing the cross-entropy loss, using negative sampling. The loss function is:
\[
\mathcal{L} = - \left( \sum_{(u, v) \in \mathcal{E}_{p,d}} \log \sigma \left( \mathbf{h}_u^\top \mathbf{h}_v \right) + \sum_{(u', v') \notin \mathcal{E}_{p,d}} \log \sigma \left( - \mathbf{h}_{u'}^\top \mathbf{h}_{v'} \right) \right),
\]
where $\sigma$ is the sigmoid function, \(\mathbf{h}_u\) and \( \mathbf{h}_v\) are the embedding representations for publication $u$ and dataset $v$, \(\mathcal{E}_{\text{p,d}}\) denotes the set of positive edges between publications and datasets.
For each positive edge \((u, v){\in}\mathcal{E}_{p,d}\), we sample uniformly at random a negative pair \((u', v'){\notin}\mathcal{E}_{p,d}\),
of publications and datasets not connected.

\section{Experimental setup}
\label{sec:evaluation}

Inspired by the evaluation conducted in \cite{irrera_etal-2025}, which systematically considers realistic conditions of increasing complexity, we evaluate \ac{MM-SAN} under three learning scenarios of increasing difficulty, distinguished by their level of inductive generalization -- i.e., by how much the model must handle nodes that were unseen during training in the \ac{SKG}.
In the \textbf{transductive} setting, all publication and dataset nodes are observed during training, the \ac{SKG} is static, and the model can exploit the full graph structure and all relationships when generating recommendations. This corresponds to the standard setup adopted in prior work~\cite{zhang_etal-2019,fu_etal-2020,mao2023hinormer}.
In the \textbf{semi-inductive} setting, new publications appear only at test time, while all datasets are already known during training. The task is to rank the existing datasets according to their relevance for these unseen publications. Note that this scenario is often referred to as “inductive’’ in the literature~\cite{zhang_etal-2019}.
Finally, in the \textbf{fully inductive} setting, both new publications and new datasets are introduced during testing. This setting mirrors realistic  conditions in which the \ac{SKG} is continuously updated with new research outputs and datasets, and recommendations must be computed for items that were not part of the training graph.

In addition, we assess the robustness of the models to missing textual metadata by considering two metadata configurations for each learning setting.
The \textbf{ideal metadata setup} represents a best-case scenario in which all datasets are fully documented, i.e., every dataset is associated with both a title and a description.
The \textbf{real metadata setup} captures more realistic situations where only 75\%, 50\%, or 25\% of the datasets have complete metadata, where the datasets with complete metadata are selected at random. This setup reflects typical \acp{SKG}, in which datasets may or may not be accompanied by rich, descriptive metadata.
The implementation of \ac{MM-SAN} and the related datasets are available at: \url{https://github.com/OrnellaIrrera/MM-SAN}.

\paragraph{\textbf{Datasets.}} We employ two \acp{SKG} extracted from the \ac{OAG}~\cite{manghi_etal-2019} and used in previous works \cite{irrera_etal-2023,irrera_etal-2025}: MES~\cite{irrera_etal-2023} and PubMed\footnote{MES and PubMed datasets are available at \url{https://doi.org/10.6084/m9.figshare.26661382.v1}}.
MES is a curated scholarly graph linking publications, datasets, and authors, and offers a reliable snapshot of the scientific activity of the \ac{MES} community within OpenAIRE.
PubMed, instead, is a larger subgraph of OAG that connects publications and datasets, as well as authors, venues, organizations, and keywords.
In contrast to MES, PubMed is not curated. \textcolor{black}{The two graphs are extremely sparse, with PubMed (99.99\%) being sparser than MES (99.97\%).}

\begin{table}[tb]
\caption{Nodes and edges statistics of the used datasets \cite{irrera_etal-2025}. $\mathcal{E}_{source,target}$ indicates the set of edges interconnecting the \textit{source} and the \textit{target}. Sparsity is computed over the whole graph as $1 - 2|\mathcal{E}| / (|\mathcal{V}|(|\mathcal{V}|-1))$, where $|\mathcal{V}|$ and $|\mathcal{E}|$ denote the total number of nodes and edges, respectively; higher values indicate higher sparsity.}
\scriptsize
\label{table:datasets}
\vspace*{-6pt}
\resizebox{0.7\textwidth}{!}{
\begin{tabular}{ll|lr|lrlr}
\toprule
Dataset & Sparsity & \multicolumn{2}{l|}{Nodes} & \multicolumn{4}{l}{Edges} \\
\midrule
\multirow{5}{*}{MES} & \multirow{5}{*}{$99.97\%$}
 & Publications (p) & \multicolumn{1}{r|}{2.1K} & $\mathcal{E}_{p,p}$ & \multicolumn{1}{r}{450}  & $\mathcal{E}_{p,a}$  & \multicolumn{1}{r}{9.8K} \\
 & & Datasets (d)     & \multicolumn{1}{r|}{3K}   & $\mathcal{E}_{d,d}$ & \multicolumn{1}{r}{1.2K} & $\mathcal{E}_{d,a}$  & \multicolumn{1}{r}{10.2K} \\
 & & Authors (a)      & \multicolumn{1}{r|}{9.5K} & $\mathcal{E}_{p,d}$ & \multicolumn{1}{r}{2.5K} &                      & \\
 & & & & & & & \\
 & & & & & & & \\
\midrule\midrule
\multirow{6}{*}{PubMed} & \multirow{6}{*}{$99.99\%$}
 & Publications (p)  & \multicolumn{1}{r|}{42.6K} & $\mathcal{E}_{p,p}$ & \multicolumn{1}{r}{18.3K} & $\mathcal{E}_{p,k}$ & \multicolumn{1}{r}{6.6K} \\
 & & Datasets (d)      & \multicolumn{1}{r|}{33.8K} & $\mathcal{E}_{d,d}$ & \multicolumn{1}{r}{8K}    & $\mathcal{E}_{d,k}$ & \multicolumn{1}{r}{17K} \\
 & & Authors (a)       & \multicolumn{1}{r|}{334K}  & $\mathcal{E}_{p,d}$ & \multicolumn{1}{r}{30.5K} & $\mathcal{E}_{p,v}$ & \multicolumn{1}{r}{42.6K} \\
 & & Venues (v)        & \multicolumn{1}{r|}{6.7K}  & $\mathcal{E}_{p,a}$ & \multicolumn{1}{r}{180K}  & $\mathcal{E}_{p,o}$ & \multicolumn{1}{r}{87.5K} \\
 & & Organizations (o) & \multicolumn{1}{r|}{14.8K} & $\mathcal{E}_{d,a}$ & \multicolumn{1}{r}{94.6K} & $\mathcal{E}_{d,o}$ & \multicolumn{1}{r}{515} \\
 & & Keywords (k)      & \multicolumn{1}{r|}{10.8K} & & & & \\
\bottomrule
\end{tabular}}
\end{table}

The statistics of the \acp{SKG} are presented in Table~\ref{table:datasets} \cite{irrera_etal-2025}, \textcolor{black}{where we report the sparsity of the two \acp{SKG}, along with the number of nodes and edges grouped by type.}

We split the set of publication--dataset edges, $\mathcal{E}_{p,d}$, into five disjoint subsets: a training set, a validation set, and three test sets corresponding to the transductive, semi-inductive, and fully inductive scenarios.

\textcolor{black}{To construct the evaluation splits, we start from the complete set of publication--dataset edges. Approximately 10\% of these edges are randomly sampled to form the fully inductive test set and a further 10\% to form the semi-inductive test set. The two sets are generated so that they involve disjoint collections of publications and datasets, ensuring that no node appears in both inductive test sets.
Once the inductive test edges have been selected, they are removed from the original graph. The inductive constraints are then enforced by removing the corresponding nodes from the remaining graph. Specifically, for the semi-inductive setting, all publications involved in the sampled test edges are removed, whereas for the fully inductive setting both the sampled publications and datasets are removed. Node removal is applied to the entire heterogeneous graph, including all incident relations (e.g., publication--author and dataset--author edges), thereby ensuring that no information about the held-out nodes is available during training and preventing information leakage.}
\textcolor{black}{The validation and transductive test sets are subsequently sampled from the remaining publication--dataset edges. Their sizes are chosen to match those of the semi-inductive and fully inductive test sets, respectively, while all residual publication--dataset edges are assigned to the training set. Finally, we verify that every publication and dataset appearing in either the validation or transductive test set is represented by at least one edge in the training graph.}

On MES, this procedure yields $2\text{K}$ edges for training, $155$ for validation, and $155$ for each of the three test sets. On PubMed, we obtain $26.8\text{K}$ training edges, $1.7\text{K}$ validation edges, and $1.7\text{K}$ edges for each test set. 

For evaluation, we treat the publications appearing in the test-set edges $\mathcal{E}_{p,d}$ as queries and aim to retrieve their associated datasets. The same \acp{SKG} were also used in previous work on dataset recommendation~\cite{irrera_etal-2024}. 
We report standard recommendation metrics -- Recall, mean reciprocal rank (MRR), and nDCG -- computed at cut-offs $k \in \{5,10\}$. These small values of $k$ reflect the fact that real-world \acp{SKG} typically associate each publication with only a limited number of datasets~\cite{irrera_etal-2025,irrera_etal-2024}. Recall measures the fraction of relevant datasets retrieved within the top-$k$ positions, nDCG evaluates ranking quality by accounting for both relevance and position, and MRR captures how early the first relevant dataset appears in the ranked list, assigning higher scores when it is ranked near the top.

\paragraph{\textbf{Parameters of MM-SAN}}
We represent each publication and dataset node with semantic vectors obtained from the pre-trained sentence-transformer \texttt{all-MiniLM-L6-v2}.
Short textual labels -- namely keywords, entities, and topics, which typically consist of one to five tokens -- are instead encoded using \texttt{Phrase-BERT}.
We derive structural representations for all nodes by learning topological embeddings with \texttt{node2vec}~\cite{grover_etal-2016}.

To enrich the graph with entity information, we run DBpedia Spotlight~\cite{mendes_etal-2011} over the textual metadata of publications and datasets, using a confidence threshold of $0.75$ so as to keep only high-precision matches and thus minimize noisy links in the \ac{SKG}. Topics are discovered with BERTopic~\cite{grootendorst_2022}, enforcing a minimum cluster size of two documents, and fine-tuned using KeyBERT to obtain representative textual labels.

\ac{MM-SAN} is trained for up to $100$ epochs with early stopping, using a learning rate of $10^{-5}$ and a mini-batch size of $1024$. For each target node, we sample five neighbors for each group of nodes, namely: five of type publication or dataset, five of type keyword/topic/entity, and five of type venue/author/organization. The cross-attention module employs $8$ attention heads and produces embeddings of dimension $128$. The training hyperparameters -- number of epochs, learning rate, batch size, number of attention heads, and per-type neighborhood size -- are selected via grid search.

\textcolor{black}{For model training, we consider four configurations corresponding to 100\%, 75\%, 50\%, and 25\% of available dataset metadata. Across these configurations, the underlying training graph is kept fixed, while the metadata associated with dataset nodes is progressively masked according to the desired availability level. A separate model is trained for each configuration, enabling the model to learn and be evaluated under consistent metadata conditions.
Each model is subsequently evaluated under all three recommendation regimes: transductive, semi-inductive and fully inductive. At test time, for the semi-inductive and fully inductive settings, the held-out nodes are reintroduced into the graph together with their non-target incident edges (e.g., publication–author, dataset–venue, publication–publication relations), while the publication–dataset edges under evaluation remain hidden from the sampling procedure to prevent information leakage.}
\textcolor{black}{For each configuration with incomplete metadata (from 75\% to 25\% availability), datasets lacking textual metadata are assigned a fixed all-ones vector in place of their textual embedding. This effectively suppresses the textual modality for those datasets while preserving the input structure required by the cross-attention mechanism, allowing the model to continue processing both modalities in a consistent manner.
}
\textcolor{black}{Overall, the experimental design independently controls two factors: \textit{metadata availability} during training and \textit{node novelty} at evaluation time. Metadata availability is varied by training separate models under different metadata-masking configurations while keeping the graph structure fixed. Conversely, node novelty is varied only at evaluation time by assessing the same trained model under transductive, semi-inductive, and fully inductive conditions. This design allows us to disentangle the impact of missing metadata from the challenges introduced by inductive generalization.}

\paragraph{\textbf{Baselines.}}
\begin{table}[t]
\footnotesize
\centering
\caption{Overview of the baselines used for comparison with MM-SAN. For each baseline, we report: name, the type (transductive or inductive), task (general, or dataset recommendation), and a brief description.}
\label{tab:baselines_overview}
\resizebox{\textwidth}{!}{
\begin{tabular}{l c c p{8cm}}
\toprule
\textbf{Model} & \textbf{Type} & \textbf{Task} & \textbf{Description} \\
\midrule
GraphSAGE \cite{hamilton_etal-2017} & Inductive & General &
Learns node embeddings by sampling and aggregating information from a node's local neighbors, enabling inductive learning on unseen nodes. \\

GAT \cite{velivckovic_etal-2017} & Inductive & General &
Employs an attention mechanism to assign different weights to neighboring nodes during message passing. \\

KGAT \cite{wang_etal-2019kgat} & Inductive & General &
Combines the user-item bipartite graph with a knowledge graph into a unified collaborative knowledge graph, applying attentive propagation with relation-aware weights. \\

HAN \cite{wang_etal-2022han} & Inductive & General &
Applies hierarchical attention -- at the node level and across semantic types of relations -- to capture diverse relational patterns in multi-type networks. \\

HGT \cite{hu_etal-2020} & Inductive & General &
Transformer-based model that incorporates type-specific parameters and relational attention to model complex interactions across node and edge types. \\

HetGNN (HGNN) \cite{zhang_etal-2019} & Inductive & General &
Sampling-and-aggregation architecture integrating random-walk--based sampling with bi-LSTM and self-attention aggregation. \\

\midrule

ST-T & Transductive/Semi & Dataset recsys &
Text-based baseline computed as the dot product between pre-trained sentence-transformer embeddings of a publication and a dataset \(
\text{ST-T}(u,v) = \mathbf{v}_u \cdot \mathbf{v}_v
\)
where $\mathbf{v}_u$ and $\mathbf{v}_v$ are the embeddings of the publication and dataset respectively. \\\midrule
KGCN \cite{wang_etal-2019kgcn} & Transductive & General &
Extends GCNs to knowledge graphs for recommendation, aggregating neighborhood information with biased weights learned from user-specific relation scores. \\

MetaPath2Vec (MP2Vec) \cite{dong_etal-2017} & Transductive & General &
Skip-gram-based representation learning method that performs metapath-guided random walks over heterogeneous networks. \\

RGCN \cite{chen2019rgcn} & Transductive & General &
Relational extension of GCNs that explicitly models different edge and node types through relation-specific transformation matrices. \\

VGAE \cite{kipf_etal-2016} & Transductive & General &
Variational graph autoencoder that learns latent node embeddings by reconstructing the adjacency matrix, without exploiting node or edge type information. \\

HVGAE \cite{altaf_etal-2019} & Transductive & Dataset recsys &
Heterogeneous graph variational autoencoder tailored for dataset recommendation, extending VGAE to multi-type scholarly networks. \\

LinearSVM (LSVM) \cite{fabrer_etal-2021} & Transductive & Dataset recsys &
Text-based baseline that recommends datasets relying on lengthy textual descriptions, such as publication abstracts, without exploiting graph structure. \\

DataFinder (DF) \cite{viswanathan-etal-2023-datafinder} & Transductive & Dataset recsys &
Neural bi-encoder model that recommends datasets given a set of publication abstracts, matching textual queries to dataset descriptions. \\

EnsembleCN (Ens) \cite{wang_etal-2022_b} & Transductive & Dataset recsys &
Ensemble method that recommends scientific datasets by exploiting author and citation networks. \\

\bottomrule
\end{tabular}}
\end{table}

We compare \ac{MM-SAN} against: (i) six general-purpose inductive graph-based baselines (also used in \cite{zhang_etal-2019,wang_etal-2022han,hu_etal-2020,irrera_etal-2025}), (ii) eight transductive baselines, and (iii) one text-based baseline. We report an overview of the baselines in Table \ref{tab:baselines_overview}, where, for each baseline, we report the learning setting it operates in (transductive or inductive), the task it was originally designed for (general-purpose graph representation learning or dataset recommendation), and a brief description of its underlying approach. 

For each baseline method, we report the best-performing configuration selected via grid search. The hyperparameter ranges explored for each baseline are summarized in the appendix. 
It is important to note that most of these baselines were not originally conceived for the dataset recommendation problem. Instead, they are general-purpose graph representation models designed to operate on attributed graphs. Evaluating \ac{MM-SAN} against such widely used, domain-agnostic approaches allows us to more clearly position its contribution within the broader \ac{GRL} landscape.

For all models, node features are initialized by concatenating textual embeddings from the pre-trained Sentence Transformer \texttt{all-MiniLM-L6-v2} with structural embeddings learned via \texttt{node2vec}.
GraphSAGE~\cite{hamilton_etal-2017} (SAGE) and GAT~\cite{velivckovic_etal-2017} are trained for $200$ epochs using a learning rate of $10^{-5}$ and early stopping, with an embedding size of $128$.
Similarly, HAN~\cite{wang_etal-2022han} and HGT~\cite{hu_etal-2020} are optimized for $200$ epochs with a learning rate of $10^{-4}$, also with early stopping and a $128$-dimensional output space.
HetGNN~\cite{zhang_etal-2019} (HGNN) is trained for $100$ epochs with early stopping, a learning rate of $10^{-5}$, and $10$ sampled neighbors per node type, following the original setup.
\textcolor{black}{KGAT was trained for up to $150$ epochs with early stopping, using a learning rate of $10^{-4}$, $2$ layers, and an embedding dimension of $128$.}
\textcolor{black}{For what concerns the transductive methods, RGCN was trained for up to $150$ epochs with early stopping, using a learning rate of $3\times10^{-4}$, $2$ layers, a hidden dimension of $128$, and a dropout rate of $0.10$.
KGCN followed the same training budget ($150$ epochs, patience $20$), with a learning rate of $1\times10^{-4}$, $2$ layers, a hidden dimension of $128$, dropout of $0.10$.
VGAE was optimized for up to $150$ epochs with early stopping, a learning rate of $5\times10^{-3}$, $2$ relational graph convolutional layers with $16$ bases, a latent dimension of $128$, and a KL-divergence term linearly warmed up over $20$ epochs to a final weight of $0.001$.
HVGAE shared the same training schedule ($150$ epochs, patience $20$), using a learning rate of $1\times10^{-3}$, $2$ layers, a hidden dimension of $128$ projected onto a $16$-dimensional latent space, and an analogous KL warm-up ($30$ epochs, final weight of $0.001$) applied independently to the publication and dataset encoders.}
Across all the selected models, negative sampling used $5$ negatives per positive edge, and parameters were optimized with AdamW using a weight decay of $1\times10^{-5}$.
All baselines have been implemented using PyTorch Geometric~\cite{fey2019fast}.



For the text-only baseline, we represent publications and datasets using embeddings derived from the pre-trained Sentence Transformer \texttt{all-MiniLM-L6-v2}.
This baseline can be instantiated only when (i) all datasets are present at training time and (ii) every dataset is associated with full textual metadata. These conditions are met in the transductive and semi-inductive scenarios with complete metadata, since ranking scores can be computed only if the entire candidate set of datasets and their textual descriptions are available.

Several existing methods are not included in our comparison.
GATNE~\cite{cen_etal-2019} is omitted because it is tailored to settings with multiple edge types over a single set of nodes and relies on user-specific information, whereas our task involves recommending datasets to publications rather than to users.
MAGNN~\cite{fu_etal-2020} is excluded as it was originally developed for node classification and does not naturally extend to inductive evaluation protocols.
Similarly, HiNormer~\cite{mao2023hinormer} is not considered, since it is not designed for recommendation tasks.

\section{Results}
\label{sec:results}

This section presents the results separately for the inductive and transductive baselines. To assess the statistical significance of our results, we conducted paired t-tests ($\alpha = 0.05$) on a per-query basis, comparing MM-SAN against each baseline. All differences were statistically significant ($p < 0.05$).

In Appendix~\ref{app:authors}, we further extend our analysis by evaluating MM-SAN on dataset recommendation for authors, a more challenging variant of the task in which authors, rather than publications, serve as query nodes.

\paragraph{\textbf{Inductive results.}} 
\begin{table*}[t]
\centering
\caption{Recall@$k$ (R@$k$), nDCG@$k$ (N@$k$) and Mean Reciprocal Rank@$k$ (MRR@$k$), with $k \in \{5,10\}$, on MES in transductive (Tran), semi-inductive (Semi), and inductive (Ind) settings, and with \(\{100,75,50,25\}\% \) of datasets with available metadata. Best results are in \textbf{bold}, second-best are \underline{underlined}.}
\label{table:mes}
\resizebox{\textwidth}{!}{
\begin{tabular}{lll|rrrrrrr|r||rrrrrrr|r}
\toprule
 & & &
\multicolumn{8}{c|}{\textbf{MES $k=5$}} &
\multicolumn{8}{c}{\textbf{MES $k=10$}} \\
\text{Metadata \%} & \text{Setting} & \text{Metric} &
GAT & KGAT & SAGE & HGT & HAN & HGNN & ST-T & MM-SAN &
GAT & KGAT & SAGE & HGT & HAN & HGNN & ST-T & MM-SAN \\
\midrule

\multirow{9}{*}{100\%} & \multirow{3}{*}{Tran}
& R@$k$ & 0.625 & 0.136 & 0.489 & 0.023 & 0.010 & 0.581 & \underline{0.693} & \textbf{0.727} &
\underline{0.748} & 0.211 & 0.571 & 0.037 & 0.019 & 0.656 & 0.721 & \textbf{0.795} \\
& & N@$k$ & 0.475 & 0.089 & 0.398 & 0.020 & 0.009 & 0.438 & \textbf{0.651} & \underline{0.546} &
0.515 & 0.113 & 0.424 & 0.030 & 0.011 & 0.455 & \textbf{0.661} & \underline{0.565} \\
& & MRR@$k$ & 0.462 & 0.073 & 0.448 & 0.020 & 0.009 & 0.378 & \underline{0.470} & \textbf{0.484} &
\underline{0.503} & 0.083 & 0.469 & 0.031 & 0.009 & 0.392 & 0.493 & \textbf{0.515} \\
\cline{2-19}

& \multirow{3}{*}{Semi}
& R@$k$ & 0.621 & 0.106 & 0.446 & 0.007 & 0.000 & 0.587 & \underline{0.696} & \textbf{0.719} &
0.719 & 0.129 & 0.522 & 0.027 & 0.006 & 0.655 & \underline{0.753} & \textbf{0.799} \\
& & N@$k$ & 0.479 & 0.066 & 0.316 & 0.003 & 0.000 & 0.417 & \textbf{0.644} & \underline{0.524} &
0.484 & 0.074 & 0.329 & 0.010 & 0.004 & 0.438 & \textbf{0.663} & \underline{0.551} \\
& & MRR@$k$ & \underline{0.454} & 0.053 & 0.437 & 0.006 & 0.000 & 0.362 & 0.450 & \textbf{0.461} &
\text{0.463} & 0.057 & \underline{0.485}& 0.007 & 0.005 & 0.370 & \text{0.472} & \textbf{0.499} \\
\cline{2-19}

& \multirow{3}{*}{Ind}
& R@$k$ & \underline{0.590} & 0.072 & 0.439 & 0.000 & 0.000 & 0.582 & - & \textbf{0.687} &
\underline{0.707} & 0.112 & 0.509 & 0.000 & 0.000 & 0.657 & - & \textbf{0.743} \\
& & N@$k$ & \underline{0.442} & 0.048 & 0.315 & 0.000 & 0.000 & 0.422 & - & \textbf{0.519} &
\underline{0.506} & 0.062 & 0.341 & 0.000 & 0.000 & 0.447 & - & \textbf{0.537} \\
& & MRR@$k$ & \underline{0.434} & 0.041 & \text{0.417} & 0.000 & 0.000 & 0.370 & - & \textbf{0.466} &
\underline{0.449} & 0.047 & 0.447 & 0.000 & 0.000 & 0.379 & - & \textbf{0.492} \\

\midrule\midrule

\multirow{9}{*}{75\%} & \multirow{3}{*}{Tran}
& R@$k$ & \underline{0.481} & 0.020 & 0.452 & 0.022 & 0.008 & 0.473 & - & \textbf{0.623} &
0.560 & 0.034 & 0.527 & 0.032 & 0.010 & \underline{0.578} & - & \textbf{0.759} \\
& & N@$k$ & \underline{0.370} & 0.014 & 0.294 & 0.011 & 0.006 & 0.356 & - & \textbf{0.447} &
0.396 & 0.019 & 0.317 & 0.021 & 0.007 & \underline{0.400} & - & \textbf{0.493} \\
& & MRR@$k$ & 0.391 & 0.012 & \underline{0.421} & 0.014 & 0.003 & 0.300 & - & \textbf{0.495} &
0.415 & 0.014 & \underline{0.464} & 0.019 & 0.007 & 0.349 & - & \textbf{0.508} \\
\cline{2-19}

& \multirow{3}{*}{Semi}
& R@$k$ & 0.483 & 0.030 & 0.432 & 0.000 & 0.000 & \underline{0.515} & - & \textbf{0.598} &
0.536 & 0.061 & 0.491 & 0.018 & 0.001 & \underline{0.571} & - & \textbf{0.704} \\
& & N@$k$ & \underline{0.404} & 0.018 & 0.247 & 0.000 & 0.000 & 0.373 & - & \textbf{0.468} &
\underline{0.421} & 0.029 & 0.272 & 0.006 & 0.001 & 0.409 & - & \textbf{0.478} \\
& & MRR@$k$ & 0.383 & 0.015 & \underline{0.398} & 0.000 & 0.000 & 0.283 & - & \textbf{0.428} &
\underline{0.409} & 0.019 & 0.408 & 0.010 & 0.000 & 0.295 & - & \textbf{0.481} \\
\cline{2-19}

& \multirow{3}{*}{Ind}
& R@$k$ & 0.409 & 0.026 & 0.385 & 0.000 & 0.000 & \underline{0.423} & - & \textbf{0.598} &
0.496 & 0.046 & 0.416 & 0.000 & 0.000 & \underline{0.541} & - & \textbf{0.726} \\
& & N@$k$ & 0.324 & 0.015 & 0.235 & 0.000 & 0.000 & \underline{0.383} & - & \textbf{0.441} &
0.354 & 0.022 & 0.245 & 0.000 & 0.000 & \underline{0.408} & - & \textbf{0.511} \\
& & MRR@$k$ & 0.313 & 0.012 & \underline{0.339} & 0.000 & 0.000 & 0.336 & - & \textbf{0.398} &
0.344 & 0.014 & \underline{0.351} & 0.000 & 0.000 & 0.339 & - & \textbf{0.403} \\

\midrule\midrule

\multirow{9}{*}{50\%} & \multirow{3}{*}{Tran}
& R@$k$ & 0.361 & 0.014 & 0.286 & 0.015 & 0.007 & \underline{0.443} & - & \textbf{0.539} &
0.417 & 0.020 & 0.338 & 0.022 & 0.008 & \underline{0.560} & - & \textbf{0.667} \\
& & N@$k$ & 0.319 & 0.005 & 0.212 & 0.012 & 0.006 & \underline{0.324} & - & \textbf{0.398} &
0.334 & 0.008 & 0.288 & 0.021 & 0.008 & \underline{0.377} & - & \textbf{0.418} \\
& & MRR@$k$ & \underline{0.311} & 0.003 & 0.237 & 0.008 & 0.005 & 0.276 & - & \textbf{0.340} &
\underline{0.343} & 0.004 & 0.306 & 0.012 & 0.006 & 0.313 & - & \textbf{0.404} \\
\cline{2-19}

& \multirow{3}{*}{Semi}
& R@$k$ & 0.349 & 0.008 & 0.240 & 0.000 & 0.000 & \underline{0.484} & - & \textbf{0.518} &
0.396 & 0.015 & 0.294 & 0.010 & 0.003 & \underline{0.519} & - & \textbf{0.700} \\
& & N@$k$ & 0.277 & 0.005 & 0.209 & 0.000 & 0.000 & \underline{0.352} & - & \textbf{0.359} &
0.295 & 0.007 & 0.226 & 0.004 & 0.001 & \underline{0.367} & - & \textbf{0.441} \\
& & MRR@$k$ & \underline{0.302} & 0.004 & 0.203 & 0.000 & 0.000 & 0.253 & - & \textbf{0.323} &
0.313 & 0.005 & 0.244 & 0.002 & 0.002 & \underline{0.325} & - & \textbf{0.382} \\
\cline{2-19}

& \multirow{3}{*}{Ind}
& R@$k$ & 0.321 & 0.007 & 0.196 & 0.000 & 0.000 & \underline{0.434} & - & \textbf{0.526} &
0.371 & 0.013 & 0.234 & 0.000 & 0.000 & \underline{0.482} & - & \textbf{0.598} \\
& & N@$k$ & 0.258 & 0.004 & 0.150 & 0.000 & 0.000 & \underline{0.339} & - & \textbf{0.391} &
0.274 & 0.006 & 0.166 & 0.000 & 0.000 & \underline{0.357} & - & \textbf{0.415} \\
& & MRR@$k$ & 0.273 & 0.003 & 0.181 & 0.000 & 0.000 & \underline{0.304} & - & \textbf{0.326} &
0.298 & 0.004 & 0.221 & 0.000 & 0.000 & \underline{0.323} & - & \textbf{0.345} \\

\midrule\midrule

\multirow{9}{*}{25\%} & \multirow{3}{*}{Tran}
& R@$k$ & 0.230 & 0.004 & 0.190 & 0.007 & 0.003 & \underline{0.336} & - & \textbf{0.467} &
0.256 & 0.020 & 0.210 & 0.015 & 0.005 & \underline{0.480} & - & \textbf{0.599} \\
& & N@$k$ & 0.203 & 0.002 & 0.165 & 0.007 & 0.002 & \underline{0.242} & - & \textbf{0.329} &
0.212 & 0.012 & 0.171 & 0.009 & 0.004 & \underline{0.265} & - & \textbf{0.372} \\
& & MRR@$k$ & \underline{0.210} & 0.002 & 0.160 & 0.005 & 0.003 & 0.169 & - & \textbf{0.283} &
\underline{0.239} & 0.009 & 0.189 & 0.007 & 0.004 & 0.181 & - & \textbf{0.340} \\
\cline{2-19}

& \multirow{3}{*}{Semi}
& R@$k$ & 0.206 & 0.003 & 0.145 & 0.000 & 0.000 & \underline{0.367} & - & \textbf{0.454} &
0.213 & 0.005 & 0.172 & 0.005 & 0.002 & \underline{0.443} & - & \textbf{0.606} \\
& & N@$k$ & 0.172 & 0.001 & 0.125 & 0.000 & 0.000 & \underline{0.270} & - & \textbf{0.306} &
0.174 & 0.002 & 0.135 & 0.002 & 0.002 & \underline{0.295} & - & \textbf{0.354} \\
& & MRR@$k$ & 0.177 & 0.001 & 0.081 & 0.000 & 0.000 & \underline{0.247} & - & \textbf{0.256} &
0.180 & 0.002 & 0.107 & 0.000 & 0.001 & \underline{0.259} & - & \textbf{0.275} \\
\cline{2-19}

& \multirow{3}{*}{Ind}
& R@$k$ & 0.159 & 0.007 & 0.095 & 0.000 & 0.000 & \underline{0.322} & - & \textbf{0.453} &
0.193 & 0.008 & 0.102 & 0.000 & 0.000 & \underline{0.394} & - & \textbf{0.539} \\
& & N@$k$ & 0.128 & 0.004 & 0.069 & 0.000 & 0.000 & \underline{0.233} & - & \textbf{0.321} &
0.139 & 0.005 & 0.075 & 0.000 & 0.000 & \underline{0.257} & - & \textbf{0.348} \\
& & MRR@$k$ & 0.136 & 0.003 & 0.070 & 0.000 & 0.000 & \underline{0.194} & - & \textbf{0.241} &
0.147 & 0.004 & 0.098 & 0.000 & 0.000 & \underline{0.216} & - & \textbf{0.255} \\

\bottomrule
\end{tabular}}
\end{table*}
\begin{table*}[t]
\centering
\caption{Recall@$k$ (R@$k$), nDCG@$k$ (N@$k$) and Mean Reciprocal Rank@$k$ (MRR@$k$), with $k \in \{5,10\}$, on PubMed in transductive (Tran), semi-inductive (Semi), and inductive (Ind) settings, and with \(\{100,75,50,25\}\% \) of datasets with available metadata. Best results are in \textbf{bold}, second-best are \underline{underlined}.}
\label{table:pubmed}
\resizebox{\textwidth}{!}{
\begin{tabular}{lll|rrrrrrr|r||rrrrrrr|r}
\toprule
 & & &
\multicolumn{8}{c||}{\textbf{PubMed $k=5$}} &
\multicolumn{8}{c}{\textbf{PubMed $k=10$}} \\
\text{Metadata \%} & \text{Setting} & \text{Metric} &
GAT & KGAT & SAGE & HGT & HAN & HGNN & ST-T & MM-SAN &
GAT & KGAT & SAGE & HGT & HAN & HGNN & ST-T & MM-SAN \\
\midrule

\multirow{9}{*}{100\%} & \multirow{3}{*}{Tran}
& R@$k$ & 0.241 & 0.021 & 0.239 & 0.016 & 0.030 & 0.284 & \textbf{0.590} & \underline{0.380}
& 0.304 & 0.063 & 0.308 & 0.026 & 0.043 & 0.341 & \textbf{0.639} & \underline{0.465} \\
& & N@$k$ & 0.181 & 0.011 & 0.179 & 0.015 & 0.023 & 0.202 & \textbf{0.523} & \underline{0.272}
& 0.201 & 0.022 & 0.210 & 0.019 & 0.032 & 0.223 & \textbf{0.539} & \underline{0.297} \\
& & MRR@$k$ & 0.203 & 0.011 & 0.204 & 0.013 & 0.024 & 0.209 & \textbf{0.456} & \underline{0.234}
& 0.205 & 0.021 & 0.235 & 0.015 & 0.031 & \underline{0.287} & \textbf{0.465} & 0.241 \\
\cline{2-19}

 & \multirow{3}{*}{Semi}
& R@$k$ & 0.231 & 0.020 & 0.221 & 0.006 & 0.010 & 0.246 & \textbf{0.596} & \underline{0.373}
& 0.303 & 0.052 & 0.292 & 0.007 & 0.021 & 0.301 & \textbf{0.639} & \underline{0.441} \\
& & N@$k$ & 0.171 & 0.010 & 0.169 & 0.005 & 0.008 & 0.177 & \textbf{0.527} & \underline{0.266}
& 0.194 & 0.031 & 0.192 & 0.007 & 0.011 & 0.196 & \textbf{0.541} & \underline{0.289} \\
& & MRR@$k$ & 0.202 & 0.010 & 0.204 & 0.006 & 0.010 & 0.222 & \textbf{0.450} & \underline{0.232}
& 0.229 & 0.023 & 0.219 & 0.006 & 0.012 & \underline{0.285} & \textbf{0.462} & 0.252 \\
\cline{2-19}

 & \multirow{3}{*}{Ind}
& R@$k$ & 0.196 & 0.013 & 0.207 & 0.000 & 0.000 & \underline{0.271} & - & \textbf{0.344}
& 0.265 & 0.018 & 0.252 & 0.003 & 0.002 & \underline{0.334} & - & \textbf{0.425} \\
& & N@$k$ & 0.134 & 0.012 & 0.133 & 0.000 & 0.000 & \underline{0.199} & - & \textbf{0.246}
& 0.156 & 0.015 & 0.154 & 0.002 & 0.001 & \underline{0.220} & - & \textbf{0.272} \\
& & MRR@$k$ & 0.182 & 0.011 & \underline{0.193} & 0.000 & 0.000 & 0.187 & - & \textbf{0.214}
& 0.184 & 0.012 & \underline{0.215} & 0.002 & 0.001 & 0.193 & - & \textbf{0.225} \\

\midrule\midrule

\multirow{9}{*}{75\%} & \multirow{3}{*}{Tran}
& R@$k$ & 0.201 & 0.018 & 0.183 & 0.006 & 0.012 & \underline{0.241} & - & \textbf{0.319}
& 0.251 & 0.030 & 0.237 & 0.008 & 0.023 & \underline{0.301} & - & \textbf{0.402} \\
& & N@$k$ & 0.149 & 0.013 & 0.139 & 0.005 & 0.009 & \underline{0.171} & - & \textbf{0.227}
& 0.166 & 0.020 & 0.157 & 0.007 & 0.012 & \underline{0.191} & - & \textbf{0.254} \\
& & MRR@$k$ & \underline{0.176} & 0.011 & 0.166 & 0.005 & 0.007 & 0.149 & - & \textbf{0.197}
& \underline{0.192} & 0.015 & 0.190 & 0.007 & 0.010 & 0.156 & - & \textbf{0.209} \\
\cline{2-19}

 & \multirow{3}{*}{Semi}
& R@$k$ & 0.186 & 0.012 & 0.174 & 0.000 & 0.007 & \underline{0.237} & - & \textbf{0.319}
& 0.250 & 0.022 & 0.227 & 0.004 & 0.009 & \underline{0.293} & - & \textbf{0.381} \\
& & N@$k$ & 0.140 & 0.010 & 0.133 & 0.000 & 0.007 & \underline{0.174} & - & \textbf{0.222}
& 0.161 & 0.016 & 0.150 & 0.003 & 0.009 & \underline{0.192} & - & \textbf{0.244} \\
& & MRR@$k$ & 0.138 & 0.010 & \underline{0.168} & 0.000 & 0.006 & 0.150 & - & \textbf{0.199}
& \underline{0.187} & 0.011 & 0.174 & 0.003 & 0.007 & 0.157 & - & \textbf{0.209} \\
\cline{2-19}

 & \multirow{3}{*}{Ind}
& R@$k$ & 0.145 & 0.009 & 0.136 & 0.000 & 0.000 & \underline{0.202} & - & \textbf{0.289}
& 0.192 & 0.012 & 0.186 & 0.000 & 0.002 & \underline{0.254} & - & \textbf{0.360} \\
& & N@$k$ & 0.104 & 0.005 & 0.097 & 0.000 & 0.000 & \underline{0.145} & - & \textbf{0.205}
& 0.119 & 0.010 & 0.115 & 0.000 & 0.002 & \underline{0.162} & - & \textbf{0.228} \\
& & MRR@$k$ & 0.112 & 0.008 & 0.120 & 0.000 & 0.000 & \underline{0.135} & - & \textbf{0.183}
& 0.138 & 0.010 & 0.127 & 0.000 & 0.001 & \underline{0.141} & - & \textbf{0.192} \\

\midrule\midrule

\multirow{9}{*}{50\%} & \multirow{3}{*}{Tran}
& R@$k$ & 0.156 & 0.016 & 0.145 & 0.005 & 0.007 & \underline{0.203} & - & \textbf{0.276}
& 0.196 & 0.022 & 0.174 & 0.006 & 0.013 & \underline{0.253} & - & \textbf{0.365} \\
& & N@$k$ & 0.116 & 0.009 & 0.111 & 0.005 & 0.007 & \underline{0.140} & - & \textbf{0.194}
& 0.129 & 0.018 & 0.120 & 0.006 & 0.012 & \underline{0.156} & - & \textbf{0.223} \\
& & MRR@$k$ & 0.115 & 0.010 & 0.123 & 0.004 & 0.006 & \underline{0.131} & - & \textbf{0.166}
& \underline{0.163} & 0.016 & 0.138 & 0.005 & 0.011 & 0.150 & - & \textbf{0.196} \\
\cline{2-19}

 & \multirow{3}{*}{Semi}
& R@$k$ & 0.144 & 0.013 & 0.129 & 0.000 & 0.001 & \underline{0.201} & - & \textbf{0.259}
& 0.185 & 0.025 & 0.167 & 0.003 & 0.002 & \underline{0.257} & - & \textbf{0.342} \\
& & N@$k$ & 0.110 & 0.010 & 0.100 & 0.000 & 0.001 & \underline{0.142} & - & \textbf{0.181}
& 0.123 & 0.021 & 0.112 & 0.002 & 0.002 & \underline{0.151} & - & \textbf{0.208} \\
& & MRR@$k$ & 0.104 & 0.011 & 0.110 & 0.000 & 0.001 & \underline{0.124} & - & \textbf{0.164}
& 0.130 & 0.020 & 0.115 & 0.002 & 0.002 & \underline{0.137} & - & \textbf{0.174} \\
\cline{2-19}

 & \multirow{3}{*}{Ind}
& R@$k$ & 0.096 & 0.012 & 0.097 & 0.000 & 0.000 & \underline{0.167} & - & \textbf{0.242}
& 0.131 & 0.019 & 0.126 & 0.000 & 0.001 & \underline{0.213} & - & \textbf{0.312} \\
& & N@$k$ & 0.086 & 0.009 & 0.079 & 0.000 & 0.000 & \underline{0.120} & - & \textbf{0.169}
& 0.098 & 0.015 & 0.089 & 0.000 & 0.001 & \underline{0.135} & - & \textbf{0.192} \\
& & MRR@$k$ & 0.089 & 0.010 & 0.088 & 0.000 & 0.000 & \underline{0.117} & - & \textbf{0.121}
& 0.114 & 0.011 & 0.094 & 0.000 & 0.001 & \underline{0.124} & - & \textbf{0.155} \\

\midrule\midrule

\multirow{9}{*}{25\%} & \multirow{3}{*}{Tran}
& R@$k$ & 0.113 & 0.008 & 0.108 & 0.000 & 0.000 & \underline{0.156} & - & \textbf{0.230}
& 0.138 & 0.012 & 0.120 & 0.006 & 0.002 & \underline{0.208} & - & \textbf{0.306} \\
& & N@$k$ & 0.084 & 0.006 & 0.089 & 0.000 & 0.000 & \underline{0.108} & - & \textbf{0.155}
& 0.092 & 0.009 & 0.096 & 0.006 & 0.001 & \underline{0.122} & - & \textbf{0.180} \\
& & MRR@$k$ & 0.103 & 0.007 & 0.100 & 0.000 & 0.000 & \underline{0.122} & - & \textbf{0.162}
& 0.122 & 0.010 & 0.111 & 0.003 & 0.001 & \underline{0.137} & - & \textbf{0.195} \\
\cline{2-19}

& \multirow{3}{*}{Semi}
& R@$k$ & 0.098 & 0.005 & 0.096 & 0.000 & 0.000 & \underline{0.152} & - & \textbf{0.239}
& 0.123 & 0.009 & 0.115 & 0.002 & 0.002 & \underline{0.204} & - & \textbf{0.307} \\
& & N@$k$ & 0.077 & 0.002 & 0.075 & 0.000 & 0.000 & \underline{0.107} & - & \textbf{0.167}
& 0.085 & 0.003 & 0.081 & 0.001 & 0.002 & \underline{0.120} & - & \textbf{0.169} \\
& & MRR@$k$ & 0.085 & 0.002 & 0.095 & 0.000 & 0.000 & \underline{0.120} & - & \textbf{0.169}
& 0.103 & 0.004 & 0.105 & 0.000 & 0.002 & \underline{0.125} & - & \textbf{0.188} \\
\cline{2-19}

 & \multirow{3}{*}{Ind}
& R@$k$ & 0.053 & 0.001 & 0.057 & 0.000 & 0.000 & \underline{0.129} & - & \textbf{0.205}
& 0.065 & 0.001 & 0.073 & 0.000 & 0.000 & \underline{0.176} & - & \textbf{0.278} \\
& & N@$k$ & 0.038 & 0.001 & 0.043 & 0.000 & 0.000 & \underline{0.090} & - & \textbf{0.138}
& 0.041 & 0.001 & 0.048 & 0.000 & 0.000 & \underline{0.106} & - & \textbf{0.162} \\
& & MRR@$k$ & 0.030 & 0.001 & 0.025 & 0.000 & 0.000 & \underline{0.106} & - & \textbf{0.162}
& 0.050 & 0.001 & 0.052 & 0.000 & 0.000 & \underline{0.110} & - & \textbf{0.170} \\

\bottomrule
\end{tabular}}
\end{table*}

Tables~\ref{table:mes},~\ref{table:pubmed}
report the results on MES and PubMed, respectively, at cutoffs 5 and 10, across the three evaluation settings: transductive, semi-inductive, and fully inductive. Each condition is assessed under progressively reduced
proportions of datasets with available metadata: 100\%, 75\%, 50\%, and 25\%. Concerning the MES dataset, MM-SAN demonstrates strong performance, outperforming all graph-based baselines across all scenarios. 
Overall, the models achieve their highest scores when full metadata is available (100\% rows in Table \ref{table:mes}). However, this scenario is the least representative of real-world academic contexts. MM-SAN preserves strong performance even in the more demanding semi-inductive and fully inductive settings, where predictions must be generated for previously unseen nodes, and remains the best graph-based model in all three settings.
Moreover, when metadata availability is reduced (from 75\% to 25\% of the datasets), MM-SAN maintains the highest effectiveness across all settings and metrics, indicating greater robustness to limited metadata. When the amount of available metadata decreases, the shift from the transductive to the semi- and fully inductive settings has a much stronger negative impact on the performance of the homogeneous baselines, such as GAT and SAGE.
This trend underscores the importance of text-based features in enhancing recommendation effectiveness, reinforcing MM-SAN's adaptability in handling sparse metadata scenarios.
MM-SAN is the only model that consistently achieves Recall@5 above 0.450, Recall@10 above 0.530, and nDCG@5 and nDCG@10 above 0.300 and 0.345 respectively, even when descriptive textual metadata is available for only 25\% of the nodes. In the most challenging scenario -- the fully inductive setting with 25\% metadata -- \ac{MM-SAN} improves over the strongest baseline (HGNN) by 41\% in Recall@5, 38\% in nDCG@5, 37\% in Recall@10, and 35\% in nDCG@10.
This indicates that the \ac{MM-SAN} multimodal approach leveraging topology-based features successfully reduces the effect of missing text-based features, a condition that significantly affects the performance of other baselines. 
HGNN, which is designed explicitly for heterogeneous \acp{SKG}, emerges as the second top-performing model after MM-SAN in terms of Recall and nDCG in the fully inductive setting when 75\%, 50\%, and 25\% of metadata are available.
We observe that SAGE and GAT -- two \ac{GRL} methods designed for homogeneous graphs -- when metadata is fully available, remain competitive with, and in the case of MES even surpass, HGNN, which is specifically designed for heterogeneous attributed graphs. On the other hand, they experience a substantial performance drop as the proportion of available metadata decreases. 
This behavior can be attributed to the ability of convolution-based message passing to fully leverage the semantic richness of node representations when all feature dimensions are available. However, once the feature space becomes sparse or incomplete, these methods face a substantial performance decline. 
The extremely sparse and imbalanced heterogeneous structure of the graph prevents HAN and HGT from learning stable relation-aware representations, which results in severe performance degradation -- particularly in inductive settings where unseen nodes cannot rely on pre-learned type-specific parameters.
\textcolor{black}{The same holds for KGAT, which achieves Recall@5 higher than 0.1 only when all the metadata are available. KGAT's attentive propagation relies on relation-specific embeddings to weigh the contribution of each neighbor. Reliably estimating these relation embeddings requires a sufficient number of observed triples per relation type. In our setting, the paper–dataset relation is extremely sparse: this scarcity limits how well the corresponding relation embedding, and consequently the attention weights computed from it, can be learned.}
While the text-based ST-T baseline achieves higher nDCG in the transductive and semi-inductive setups, MM-SAN obtains higher Recall and MRR in both settings, showcasing its robustness across different conditions on the MES dataset.  

As shown in Table~\ref{table:pubmed}, MM-SAN exhibits strong robustness when applied to real-world, sparse, an large \acp{SKG} like PubMed. Similar to the MES dataset, all tested methods experience a performance decline in semi-inductive and fully inductive settings as textual metadata becomes less available -- a natural outcome given the increasing difficulty of the task. 
Compared to MES, performance on PubMed in terms of Recall, nDCG and MRR is noticeably lower. PubMed is considerably more sparse, authors are not disambiguated -- further increasing sparsity -- it is not manually curated, and its metadata is of lower quality. These factors collectively and directly degrade the overall model performance. 
MM-SAN demonstrates the highest resilience to reduced textual metadata and is the best graph-based model in all settings, with the only exception of MRR@10 with full metadata in the transductive and semi-inductive settings, where HGNN performs better. It is the only method that consistently maintains a Recall@5 above 0.200, Recall@10 above 0.275, an nDCG@5 above 0.135, and an nDCG@10 at least 0.162 across all settings.
Also, in the most challenging scenario, with only 25\% of datasets containing available metadata and within a full-inductive setup, \ac{MM-SAN} surpasses the best baseline (HGNN) by 59\% in Recall@5, 58\% in Recall@10, and 53\% in nDCG@5 and nDCG@10. 
Consistently with the results on the MES dataset, SAGE and GAT, designed for homogeneous graphs, rank among the strongest baselines after HGNN, achieving their best performance in the ideal scenario with 100\% of metadata available.
KGAT, HAN and HGT exhibit the lowest performance on the PubMed dataset, highlighting their limited effectiveness in real-world \acp{SKG}. 
The text-based baseline, ST-T, achieves the highest effectiveness in transductive and semi-inductive scenarios with full metadata availability, surpassing all graph-based methods (including MM-SAN) in Recall, nDCG and MRR. However, ST-T's effectiveness is limited to ideal conditions, making it less suited to addressing the open challenges of real-world \acp{SKG}.
\begin{table*}[t]
\centering
\scriptsize
\caption{Recall@$k$ (R@$k$), nDCG@$k$ (N@$k$), and Mean Reciprocal Rank@$k$ (MRR@$k$) with $k=5$ on MES and PubMed for the purely transductive methods and MM-SAN, with \(\{100,75,50,25\}\%\) of datasets with available metadata. Best results are in \textbf{bold}, second-best are \underline{underlined}. LSVM=LinearSVM, DF=DataFinder, Ens=EnsembleCN, MP2Vec=MetaPath2Vec. }
\label{tab:trans}
\begin{tabular}{ll|rrrrrrrr|r}
\toprule
& &
\multicolumn{9}{c}{\textbf{MES $k=5$}} \\
\text{Metadata \%} & \text{Metric} &
LSVM & DF & Ens & RGCN & KGCN & HVGAE & VGAE & MP2Vec & MM-SAN \\
\midrule
\multirow{3}{*}{100\%}
& R@$k$   & 0.095& \underline{0.420} & 0.312& 0.085 & 0.027 & 0.007 & \text{0.414} & 0.032 & \textbf{0.727} \\
& N@$k$   &0.076 & \underline{0.322}&0.129 & 0.054 & 0.009 & 0.004 & \text{0.330} & 0.021 & \textbf{0.546} \\
& MRR@$k$ & 0.073& \underline{0.310}&0.229& 0.046 & 0.009 & 0.003 & \text{0.313} & 0.018 & \textbf{0.484} \\
\midrule
\multirow{3}{*}{75\%}
& R@$k$   & - & \underline{0.325}& 0.120& 0.048 & 0.020 & 0.007 & \text{0.308} & - & \textbf{0.623} \\
& N@$k$   & -& \underline{0.312}& 0.100& 0.023 & 0.012 & 0.003 & \text{0.249} & - & \textbf{0.447} \\
& MRR@$k$ &- &\underline{0.301}& 0.140& 0.016 & 0.004 & 0.002 & \text{0.224} & - & \textbf{0.495} \\
\midrule
\multirow{3}{*}{50\%}
& R@$k$   & -& \underline{0.412}& 0.090& 0.014 & 0.014 & 0.003 & \text{0.333} & -& \textbf{0.539} \\
& N@$k$   &- & \underline{0.302}& 0.076& 0.005 & 0.005 & 0.003 & \text{0.287} & - & \textbf{0.398} \\
& MRR@$k$ & -& \underline{0.298}& 0.101& 0.002 & 0.003 & 0.002 & \text{0.250} & - & \textbf{0.340} \\
\midrule
\multirow{3}{*}{25\%}
& R@$k$   & -& \underline{0.299}& 0.010& 0.008 & 0.007 & 0.002 & \text{0.292} & - & \textbf{0.467} \\
& N@$k$   & -& 0.198& 0.008& 0.003 & 0.003 & 0.002 & \underline{0.232} & - & \textbf{0.329} \\
& MRR@$k$ & -& \underline{0.229}& 0.012& 0.006 & 0.004 & 0.001 & \text{0.218} & - & \textbf{0.283} \\
\midrule\midrule
& &
\multicolumn{9}{c}{\textbf{PubMed $k=5$}} \\
\text{Metadata \%} & \text{Metric} &
LSVM & DF & Ens & RGCN & KGCN & HVGAE & VGAE & MP2Vec & MM-SAN \\
\midrule
\multirow{3}{*}{100\%}
& R@$k$   &0.074 & \underline{0.280}&0.005 & 0.001 & 0.006 & 0.001 & 0.024 & \text{0.061} & \textbf{0.380} \\
& N@$k$   & 0.060&\underline{0.210} & 0.002& 0.001 & 0.003 & 0.001 & 0.012 & \text{0.040} & \textbf{0.272} \\
& MRR@$k$ &0.057 & \underline{0.121}& 0.005& 0.001 & 0.002 & 0.001 & 0.008 & \text{0.033} & \textbf{0.234} \\
\midrule
\multirow{3}{*}{75\%}
& R@$k$   & -&\underline{0.129} &0.003 & 0.001 & 0.001 & 0.000 & 0.029 & - & \textbf{0.319} \\
& N@$k$   & -&\underline{0.108} &0.001 & 0.000 & 0.001 & 0.000 & 0.015 & - & \textbf{0.227} \\
& MRR@$k$ & -&\underline{0.103} &0.003 & 0.000 & 0.000 & 0.000 & 0.010 & - & \textbf{0.197} \\
\midrule
\multirow{3}{*}{50\%}
& R@$k$   & -&\underline{0.109}&0.001 & 0.001 & 0.001 & 0.000 & 0.018 & - & \textbf{0.276} \\
& N@$k$   & -&\underline{0.090}&0.001& 0.001 & 0.003 & 0.000 & 0.009 & - & \textbf{0.194} \\
& MRR@$k$ & -&\underline{0.080} &0.002 & 0.001 & 0.000 & 0.000 & 0.007 & - & \textbf{0.166} \\
\midrule
\multirow{3}{*}{25\%}
& R@$k$   & -&\underline{0.092} &0.000 & 0.001 & 0.000 & 0.000 & 0.012 & - & \textbf{0.230} \\
& N@$k$   & -&\underline{0.072} &0.000 & 0.000 & 0.000 & 0.000 & 0.007 & - & \textbf{0.155} \\
& MRR@$k$ & -&\underline{0.078} &0.000 & 0.000 & 0.000 & 0.000 & 0.005 & - & \textbf{0.162} \\
\bottomrule
\end{tabular}
\end{table*}

\paragraph{\textcolor{black}{\textbf{Transductive results.}}} In Table~\ref{tab:trans}, we compare the performance of MM-SAN against a set of baselines specifically designed for purely transductive recommendation. We report the results on the MES and PubMed datasets at a cutoff of 5 under the transductive setup, considering progressively decreasing proportions of available metadata.
\text{MetaPath2Vec and LinearSVM} are reported only at 100\% metadata, as neither uses dataset textual descriptions: MetaPath2Vec relies purely on graph structure, and LinearSVM only on publication text. Their results are therefore unaffected by dataset metadata removal.
\text{RGCN, KGCN, and HVGAE} all perform poorly (R@5~$<$~0.1 on MES, $<$~0.01 on PubMed). RGCN's relation-specific weights cannot be reliably learned from so few publication--dataset edges (2K on MES, 27K on PubMed). KGCN's neighbor weighting depends on a \text{user-specific} interaction history, but each publication-query is linked to fewer than one dataset on average; the model thus degenerates to near-uniform aggregation. 
HVGAE's dataset encoder relies solely on the (mostly empty) bipartite row of paper–dataset links, with no textual features to fall back on, so it cannot distinguish datasets under such extreme sparsity. Its GraphSAGE paper encoder finds almost no neighbors to aggregate on MES or PubMed and effectively collapses to encoding each paper in isolation.
\text{EnsembleCN (Ens)} performs moderately on MES (R@5~=~0.312 at 100\%, 0.010 at 25\%) but fails on PubMed (R@5~$\leq$~0.005). It relies on co-author and citation network exploration, which is only viable on MES, where authors are disambiguated and consequently well-connected; on PubMed, undisambiguated authors vastly outnumber papers, fragmenting the co-authorship network.
\text{VGAE} follows the same pattern (R@5~=~0.414$\to$0.292 on MES; $\leq$~0.029 on PubMed). Lacking type information, VGAE treats all nodes and edges as homogeneous, so its embeddings depend purely on local neighborhood density rather than node/edge semantics. This proxy is informative enough on the smaller, curated MES graph, but breaks down on the larger, sparser PubMed graph, where local density becomes too weak and noisy to distinguish nodes meaningfully.
DataFinder (DF), a purely text-based bi-encoder, is the strongest transductive baseline at full metadata (R@5~=~0.420 on MES, 0.280 on PubMed), confirming the value of semantic information when reliable. However, it cannot exploit graph structure or generalize to unseen nodes, and degrades sharply as metadata shrinks (R@5~=~0.299 on MES, 0.092 on PubMed at 25\%), making it poorly suited to realistic, metadata-incomplete settings.

Overall, we can see that the dataset recommendation methods proposed so far (EnsembleCN, DataFinder, LinearSVM, HVGAE) often assume the availability of reliable textual metadata and highly interconnected networks, while largely overlooking the dynamic nature of real-world graphs. These assumptions limit their applicability to real-world \acp{SKG}, which are typically characterized by incomplete or noisy metadata, sparse and heterogeneous connectivity, and continuously evolving structures.

\section{\textcolor{black}{Ablation study}}
\label{sec:ablation}
\begin{table}[t]
\caption{Ablation study on \text{MES}. Results are reported for the transductive setting with 100\% metadata (Tran[100\%]) and the fully inductive setting with 50\% metadata (Ind[50\%]). Results are grouped by the pipeline phase being ablated (Augmentation, Sampling, Aggregation, Combination). The best result for each metric and phase is \underline{underlined}; the full MM-SAN model (bottom row) is shown in \textbf{bold}.}
\label{tab:ablation_mes}
\scriptsize
\begin{tabular}{ll|rrr|rrr}
\toprule
& & \multicolumn{3}{c|}{\textbf{Tran [100\%]}} & \multicolumn{3}{c}{\textbf{Ind [50\%]}} \\
\cline{3-8}
Component & Variant & R@5 & nDCG@5 & MRR@5 & R@5 & nDCG@5 & MRR@5 \\
\midrule
\multirow{3}{*}{Augmentation}
& No augmentation      & 0.642 & 0.476 & 0.450 & 0.415 & 0.310 & 0.268 \\
& No entity linking    & 0.676 & 0.501 & \text{0.469} & \text{0.463} & 0.325 & 0.289 \\
& No topic modeling    & \underline{0.708} & \underline{0.512} & \underline{0.471} & \underline{0.482} & \underline{0.349} & \underline{0.301} \\
\midrule\midrule
\multirow{3}{*}{Sampling}
& Random, 1-hop        & 0.602 & 0.434 & 0.413 & 0.398 & 0.299 & 0.278 \\
& Random, 2-hop        & 0.637 & 0.471 & 0.419 & 0.403 & 0.311 & 0.291 \\
& Random, 3-hop         & \underline{0.644} & \underline{0.475} & \underline{0.428} & \underline{0.414} & \underline{0.313} & \underline{0.297} \\
\midrule\midrule
\multirow{4}{*}{Aggregation}
& GRU                   & \underline{0.642} & \underline{0.492} & \underline{0.442} & \underline{0.519} & \underline{0.367} & \underline{0.318} \\
& LSTM                  & 0.640 & 0.455 & 0.403 & \text{0.509} & 0.360 & 0.307 \\
& Multi-head attention  & 0.615 & \text{0.416} & 0.428 & 0.495 & \text{0.347} & \text{0.311} \\
& Linear projection     & 0.520 & 0.406 & 0.371 & 0.444 & 0.339 & 0.302 \\
\midrule\midrule
\multirow{4}{*}{Combination}
& GRU                   & 0.687 & 0.485 & 0.416 & 0.431 & \underline{0.332} & \underline{0.266} \\
& LSTM                  & 0.659 & 0.425 & 0.414 & 0.415 & 0.310 & 0.242 \\
& Multi-head attention  & \underline{0.693} & \underline{0.508} & \underline{0.420} & \underline{0.446} & 0.310 & 0.264 \\
& Mean pooling          & 0.621 & 0.497 & 0.339 & 0.329 & 0.256 & 0.218 \\
\midrule\midrule
\textbf{MM-SAN } & & \textbf{0.727} & \textbf{0.546} & \textbf{0.484} & \textbf{0.526} & \textbf{0.391} & \textbf{0.326} \\
\bottomrule
\end{tabular}
\vspace{2pt}

\end{table}

\begin{table}[t]
\caption{Ablation study on \text{PubMed}. Results are reported for the transductive setting with 100\% metadata (Tran[100\%]) and the fully inductive setting with 50\% metadata (Ind[50\%]). Results are grouped by the pipeline phase being ablated (Augmentation, Sampling, Aggregation, Combination). For each metric and phase, the best result for each metric is \underline{underlined}; the full MM-SAN model (bottom row) is shown in \textbf{bold}.}
\label{tab:ablation_pubmed}
\scriptsize
\begin{tabular}{ll|rrr|rrr}
\toprule
& & \multicolumn{3}{c|}{\textbf{Tran [100\%]}} & \multicolumn{3}{c}{\textbf{Ind [50\%]}} \\
\cline{3-8}
Component & Variant & R@5 & nDCG@5 & MRR@5 & R@5 & nDCG@5 & MRR@5 \\
\midrule
\multirow{3}{*}{Augmentation}
& No augmentation      & 0.311 & 0.201 & 0.181 & 0.168 & 0.120 & 0.081 \\
& No entity linking    & \underline{0.341} & \underline{0.239} & 0.199 & \underline{0.194} & \underline{0.134} & \underline{0.110} \\
& No topic modeling    & 0.329 & 0.225 & \underline{0.203} & 0.188 & 0.123 & 0.099 \\
\midrule\midrule
\multirow{3}{*}{Sampling}
& Random, 1-hop        & \underline{0.314} & \underline{0.243} & \underline{0.220} & \underline{0.190} & 0.101 & \underline{0.099} \\
& Random, 2-hop        & 0.313 & 0.225 & 0.197 & 0.176 & 0.110 & 0.078 \\
& Random, 3-hop         & 0.295 & 0.212 & 0.184 & 0.172 & \underline{0.120} & 0.069 \\
\midrule\midrule
\multirow{4}{*}{Aggregation}
& GRU                   & 0.360 & 0.264 & 0.215 & \underline{0.220} & 0.127 & 0.102 \\
& LSTM                  & \underline{0.363} & \underline{0.266} & \underline{0.222} & 0.200 & \underline{0.154} & \underline{0.119} \\
& Multi-head attention  & 0.284 & 0.203 & 0.177 & 0.187 & 0.127 & 0.112 \\
& Linear projection     & 0.257 & 0.191 & 0.169 & 0.162 & 0.121 & 0.100 \\
\midrule\midrule
\multirow{4}{*}{Combination}
& GRU                   & 0.315 & \underline{0.216} & 0.180 & 0.188 & \underline{0.139} & 0.099 \\
& LSTM                  & 0.290 & 0.181 & 0.126 & 0.210 & 0.126 & 0.098 \\
& Multi-head attention  & \underline{0.320} & 0.192 & \underline{0.221} & \underline{0.215} & 0.127 & \underline{0.101} \\
& Mean pooling          & 0.189 & 0.123 & 0.108 & 0.192 & 0.119 & 0.082 \\
\midrule\midrule
\textbf{MM-SAN } & & \textbf{0.380} & \textbf{0.272} & \textbf{0.234} & \textbf{0.242} & \textbf{0.169} & \textbf{0.121} \\
\bottomrule
\end{tabular}
\vspace{2pt}

\end{table}

In this section, we present an ablation study on the MES and PubMed datasets. Tables~\ref{tab:ablation_mes} and~\ref{tab:ablation_pubmed} assess the contribution of each phase of the pipeline, reporting Recall, nDCG, and MRR. Table~\ref{tab:split} reports a modality ablation, in which the textual or the topological embeddings of dataset nodes are partially or completely removed, to assess the relative importance of each modality and the ability of MM-SAN to compensate for a missing one.

We consider two experimental settings: a transductive scenario with complete metadata availability (Tran 100\%) and an inductive scenario where only 50\% of the metadata is available.
To assess the statistical significance of the results, we conducted paired t-tests ($\alpha$ = 0.05) on a per-query basis to compare MM-SAN against the tested ablation configurations. All comparisons yielded statistically significant differences ($p$ $<$ 0.05). 

\paragraph{\textbf{Augmentation}} For the augmentation phase, we evaluated the performance of MM-SAN under three settings: without both entity and topic nodes, without entity nodes only, and without topic nodes only. This analysis was conducted to assess the importance of the augmentation phase and to determine which type of node contributes most significantly to the model's performance.
The results are summarized in Tables \ref{tab:ablation_mes} and \ref{tab:ablation_pubmed} in \textit{Augmentation} block.

The results show that the augmentation phase is crucial for improving graph connectivity and, consequently, enhancing the dataset recommendation task by mitigating the effects of graph sparsity. Interestingly, we observe opposite trends on the MES and PubMed datasets. On MES, removing entity nodes leads to a larger performance drop than removing topic nodes (R@5: 0.676 vs. 0.708 in the Tran 100\% setting). This can be explained by the fact that MES is a curated and relatively dense dataset, where entities provide highly informative links between publications and datasets. In contrast, topics mainly act as hub nodes, offering more generic semantic connections and therefore contributing less to the recommendation performance.

On PubMed, however, the opposite behavior emerges. While entities still provide useful information, their impact is less pronounced than that of topics. Due to the much larger size and higher sparsity of the PubMed graph, topic nodes play a more important role in connecting otherwise distant regions of the graph, resulting in a greater performance gain. This is reflected by the larger drop observed when topic nodes are removed compared to entity nodes (R@5: 0.329 vs. 0.341 in tran 100\% setting), highlighting the importance of topic-based augmentation in sparse and large-scale graphs.

\paragraph{\textbf{Sampling}} For the sampling phase, we compared the random-walk strategy used in our original pipeline with random sampling at 1-, 2-, and 3-hop distances, i.e., by randomly sampling the neighbors of a node at increasing distances from the target. Overall, random walk proves to be a more effective strategy, as it allows us to weight candidate neighbors according to both their distance from and their similarity to the initial node. This enables the model to prioritize more relevant nodes while still exploring a broader portion of the graph.
The results are provided in Tables \ref{tab:ablation_mes}, \ref{tab:ablation_pubmed} in \textit{Sampling} blocks.

The results reveal different trends across the two datasets. On MES, randomly sampling neighbors at greater hop distances leads to a slight performance improvement across all learning settings (R@5 Ind [50\%]: $0.398 \rightarrow 0.403 \rightarrow 0.414$). This is consistent with the denser and better-connected structure of MES, where exploring further hops is more likely to uncover diverse and informative nodes. In contrast, on PubMed, performance is relatively stable across different hop distances (R@5 Ind [50\%]: $0.190 \rightarrow 0.176 \rightarrow 0.172$), with 1-hop sampling providing the best performance among the random-sampling strategies. This behavior is expected given the higher sparsity and strong node-type imbalance of PubMed: exploring further hops tends to introduce a larger number of nodes of the same type that are only weakly connected to the target, providing limited additional information.
Nevertheless, the random-walk strategies integrated into our pipeline consistently outperform pure random sampling. By explicitly balancing graph proximity and textual similarity, random walks can construct more diverse and informative neighborhoods around the target nodes, while avoiding the limitations imposed by the strong structural and type imbalance of the underlying \ac{SKG}.

\paragraph{\textbf{Aggregation}} For the aggregation phase, we evaluated different strategies for combining the topological and textual information of the nodes. Cross-attention jointly models the two modalities by allowing the representation of each modality to selectively incorporate information from the other, thereby learning their interactions rather than simply combining them through a fixed transformation. We also evaluated alternative aggregation strategies, including GRU, LSTM, multi-head attention, and linear projection. The results are provided in Tables \ref{tab:ablation_mes}, \ref{tab:ablation_pubmed} in \textit{Aggregation} blocks.

Overall, cross-attention provides the best performance across both datasets and learning settings, confirming the importance of explicitly modeling the interactions between topology and textual information. On MES, the best alternative is GRU, which achieves an R@5 of 0.642 and 0.519 in the transductive and inductive settings, respectively (MM-SAN: $0.727$ and $0.526$). On PubMed, LSTM is the strongest alternative in the transductive setting, with an R@5 of 0.363 (MM-SAN: $0.380$), while GRU performs best in the inductive setting, with an R@5 of 0.220 (MM-SAN: $0.242$). The same trend is reflected in the other metrics, with MM-SAN consistently achieving the highest overall performance. These results indicate that recurrent architectures can effectively capture dependencies between the modalities, but remain less effective than the explicit cross-modal interactions learned by cross-attention.
The performance gap becomes more pronounced for simpler aggregation strategies. In particular, linear projection achieves the lowest R@5 among the alternatives in the transductive setting, with 0.520 on MES and 0.257 on PubMed, while multi-head attention achieves 0.615 and 0.284, respectively. In the inductive setting, linear projection similarly obtains lower R@5 values on both datasets (0.444 on MES and 0.162 on PubMed), compared to 0.526 and 0.242 for MM-SAN. Overall, the consistent advantage of cross-attention under both complete and incomplete metadata availability suggests that explicitly modeling cross-modal interactions is particularly beneficial for dataset recommendation, where effectively integrating structural and textual information is crucial.

\paragraph{\textbf{Combination}} For the combination phase, we compared the concatenation strategy adopted by MM-SAN with four alternative ways of merging the group-specific representations produced by the aggregation module: GRU, LSTM, multi-head attention, and mean pooling. Results are reported in the \textit{Combination} block of Tables~\ref{tab:ablation_mes} and~\ref{tab:ablation_pubmed}. Multi-head attention is the strongest alternative across both datasets and settings, reaching an R@5 of 0.693 on MES and 0.320 on PubMed in the transductive setting (MM-SAN: 0.727 and 0.380), and 0.446 and 0.215 in the inductive setting (MM-SAN: 0.526 and 0.242). GRU performs close to multi-head attention on MES in the transductive setting (R@5 = 0.687), but falls further behind in the inductive one (R@5 = 0.431 on MES and 0.188 on PubMed). Mean pooling is the weakest strategy overall: in the transductive setting its R@5 drops to 0.621 on MES and 0.189 on PubMed, and it also yields the lowest inductive score on MES (R@5 = 0.329). On PubMed in the inductive setting, instead, all four alternatives fall within a narrow range (R@5 = 0.188--0.215), well below MM-SAN. Overall, these results suggest that concatenation is more effective than applying further learned or averaging operations, likely because it preserves the complementary information carried by each group of node types rather than compressing it into a single shared representation.


\begin{table}[t]
\caption{Modality ablation analysis. Each percentage indicates the proportion of dataset nodes for which the masked modality is available (e.g., [50\%]: the modality is missing for half of the dataset nodes; [0\%]: it is missing for all of them), while the other modality remains available for all dataset nodes. \textbf{Topology unavailable} denotes the setting in which topology is masked, i.e., available only for the indicated percentage of dataset nodes, while text is always fully available; \textbf{Text unavailable} denotes the symmetric setting (the original MM-SAN configuration), in which textual metadata is available only for the indicated percentage of dataset nodes while topology is always fully available.}\label{tab:split}
\footnotesize
\centering
\begin{tabular}{ll|rr||rr}

\toprule
& & \multicolumn{2}{c||}{\textbf{MES}} & \multicolumn{2}{c}{\textbf{PubMed}} \\
\cline{3-6}
Setting & Metric
& Topology unavailable & Text unavailable
& Topology unavailable & Text unavailable\\
\midrule
\multirow{3}{*}{Tran [50\%]}
& R@5    & \textbf{0.635} & 0.539 & \textbf{0.312} & 0.276 \\
& nDCG@5 & \textbf{0.423} & 0.398 & \textbf{0.267} & 0.194 \\
& MRR@5  & \textbf{0.369} & 0.340 & \textbf{0.208} & 0.166 \\
\midrule
\multirow{3}{*}{Ind [50\%]}
& R@5    & \textbf{0.603} & 0.526 & \textbf{0.254} & 0.242 \\
& nDCG@5 & \textbf{0.401} & 0.391 & \textbf{0.219} & 0.169 \\
& MRR@5  & \textbf{0.345} & 0.326 & \textbf{0.192} & 0.121 \\
\midrule\midrule
\multirow{3}{*}{Tran [0\%]}
& R@5    & \textbf{0.446} & 0.395 & \textbf{0.232} & 0.198 \\
& nDCG@5 & \textbf{0.331} & 0.301 & \textbf{0.196} & 0.134 \\
& MRR@5  & \textbf{0.220} & 0.199 & \textbf{0.154} & 0.126 \\
\midrule
\multirow{3}{*}{Ind [0\%]}
& R@5    & \textbf{0.376} & 0.338 & \textbf{0.180} & 0.149 \\
& nDCG@5 & \textbf{0.298} & 0.276 & \textbf{0.127} & 0.108 \\
& MRR@5  & \textbf{0.262} & 0.212 & \textbf{0.158} & 0.107 \\
\bottomrule
\end{tabular}
\end{table}

\paragraph{\textbf{Modality ablation.}} We further extend our ablation study to assess the relative contribution of the topological and textual modalities and to evaluate the robustness of the cross-attention mechanism when one modality is partially or completely unavailable for the target dataset nodes. Following the missing-metadata setting described in Section~\ref{sec:evaluation}, an unavailable modality is represented by replacing its embedding with a fixed constant vector of ones, preserving the dimensionality of the original representation while removing any node-specific information from that modality. 

We consider two complementary settings. First, in the \textit{Partially missing modality} setting, one modality is made unavailable for 50\% of the dataset nodes, while the other modality remains fully available. We evaluate two symmetric variants: \textit{Topology unavailable}, where the topological embeddings of the selected dataset nodes are replaced by the constant vector while their textual embeddings are preserved, and \textit{Text unavailable}, where the textual embeddings are replaced while topology remains available. Second, in the \textit{Complete missing modality} setting, the same procedure is extended to all dataset nodes (0\% availability). In this case, either all textual embeddings or all topological embeddings are replaced by the constant vector, allowing us to isolate the contribution of each modality to the recommendation of dataset nodes.

Importantly, we remove modalities only from dataset nodes, which represent the target nodes in our recommendation task, while preserving the multimodal information available for the remaining node types. This setting allows us to specifically investigate the impact of missing information on recommendation targets without reducing the entire graph to a unimodal setting. It also provides a direct assessment of whether MM-SAN can effectively exploit either modality when the other is unavailable and whether its cross-attention mechanism remains robust when topology, rather than text, constitutes the missing modality. Table~\ref{tab:split} reports Recall@5, nDCG@5, and MRR@5 for MES and PubMed under both transductive and inductive settings.

The results confirm the bidirectional nature of MM-SAN's cross-attention mechanism: regardless of which modality is masked, the remaining one is able to partially compensate for the missing signal, allowing the model to preserve reasonably strong performance in both settings. At the same time, the two settings are not perfectly symmetric. Across both datasets and learning settings, the \textit{Topology unavailable} setting tends to yield slightly higher scores than the \textit{Text unavailable} — e.g., on MES in the transductive setting at 50\% availability, Recall@5 is 0.635 when topology is missing versus 0.539 when text is missing, with a comparable gap on PubMed (0.312 vs. 0.276). This suggests that, when forced to rely on a single modality, textual information tends to provide a stronger fallback signal than topology alone, likely because it more directly encodes the semantic relevance between publications and datasets, while topology offers a more indirect, relational form of evidence. 

\section{\textcolor{black}{Discussion}}
\label{sec:discussion}
In the following, we discuss the results with respect to each research question introduced in Section~\ref{sec:intro}, and then outline the theoretical and practical implications of MM-SAN for dataset recommendation over real-world scholarly infrastructures.

\paragraph{\textcolor{black}{\textbf{RQ1 (Robustness).}}} The first research question concerns robustness to the structural sparsity of \acp{SKG}. Our results show that MM-SAN is robust to sparsity thanks to two complementary components of its architecture. First, the augmentation phase connects otherwise unrelated regions of the graph: entity and topic nodes act as hubs that link publications and datasets sharing no direct structural path. Tables~\ref{tab:ablation_mes} and~\ref{tab:ablation_pubmed} confirm that removing augmentation, or either of its two sources, consistently degrades performance on both MES and PubMed, showing that the improvement stems from bridging distant regions of the graph rather than from merely adding nodes. The relative contribution of the two sources depends on the graph: entity nodes are more beneficial on denser and curated graphs such as MES, whereas topic nodes provide a larger benefit on sparser graphs such as PubMed, where topic-based links recover connections that the original topology lacks. Second, the weighted random-walk sampling favors neighbors that are both structurally close and semantically similar to the target, instead of spending neighbor slots on weakly informative nodes, as uniform sampling tends to do. Accordingly, it consistently outperforms uniform random sampling at every hop distance.

\begin{takeawaybox}
\textbf{Takeaway.} MM-SAN copes with the structural sparsity of \acp{SKG} by enriching the graph with entity and topic nodes, which connect otherwise distant regions, and by using weighted random walks to select the most informative neighbors within sparse and unbalanced neighborhoods.
\end{takeawaybox}

\paragraph{\textcolor{black}{\textbf{RQ2 (Flexibility).}}} Flexibility refers to the ability of MM-SAN to recommend datasets in transductive, semi-inductive, and fully inductive settings, i.e., also for nodes not seen during training. Across both datasets and all metadata-availability levels (Tables~\ref{table:mes}--\ref{table:pubmed}), all methods achieve their best results in the transductive setting, but MM-SAN shows the smallest degradation when moving to the semi- and fully inductive settings, and is the only method that remains competitive in all of them, which makes it suitable for dynamic graphs where retraining is impractical or infeasible. This behavior follows from how MM-SAN builds node representations: rather than relying on node-specific parameters learned during training, it constructs the representation of a node at inference time from its sampled neighborhood. For a previously unseen publication or dataset, the weighted random walks select the most structurally close and semantically relevant nodes in its local context, so that MM-SAN can build a meaningful representation. Since no node-specific parameters need to be learned or updated, the same trained model can be directly applied to newly introduced nodes without retraining.

\begin{takeawaybox}
\textbf{Takeaway.} MM-SAN generalizes to unseen nodes because it builds their representations at inference time from neighbors selected through weighted random walks, rather than from node-specific parameters. The same trained model can therefore serve new publications and datasets without retraining, making MM-SAN particularly suitable for dynamic scholarly graphs.
\end{takeawaybox}

\paragraph{\textcolor{black}{\textbf{RQ3 (Versatility).}}} We define versatility as the ability to limit performance degradation as textual metadata becomes progressively scarcer (100\%~$\rightarrow$~75\%~$\rightarrow$~50\%~$\rightarrow$~25\%). Tables~\ref{table:mes} and~\ref{table:pubmed} show that, while several baselines perform well when textual metadata is fully available, MM-SAN degrades considerably more gracefully as metadata decreases, since its cross-attention aggregation can shift weight toward structural information. This highlights the benefit of jointly exploiting complementary modalities rather than relying exclusively on textual information. This adaptation works in both directions: the modality ablation in Table~\ref{tab:split} shows that MM-SAN also preserves reasonable performance when topology, rather than text, is missing, by relying more heavily on the textual signal. Cross-attention therefore does not implement a fixed fallback strategy, but adjusts the contribution of each modality according to its availability and informativeness, which makes MM-SAN well suited to real-world \acp{SKG} with incomplete or unreliable metadata.

\begin{takeawaybox}
\textbf{Takeaway.} MM-SAN's versatility stems from its cross-attention mechanism, which adjusts the contribution of textual and topological information according to their availability, allowing the model to rely on whichever modality remains informative.
\end{takeawaybox}

\paragraph{\textcolor{black}{\textbf{Theoretical and practical implications of MM-SAN}}} From a theoretical perspective, our findings suggest that dataset recommendation over real-world \acp{SKG} should jointly account for the semantic and structural dimensions of scholarly resources. Since datasets are sparsely connected to publications and their metadata is often incomplete, the two modalities provide complementary evidence: text captures the semantic relationship between a publication and a dataset, while graph structure provides useful signals when descriptions are missing or of low quality. This is reflected in our experiments, where MM-SAN maintains comparatively strong performance as metadata availability decreases and in the more challenging inductive settings. More broadly, robustness to incomplete multimodal information and the ability to generalize to evolving graphs emerge as key properties for recommender systems operating on scholarly infrastructures.

From a practical perspective, MM-SAN could be integrated into scholarly repositories, digital libraries, or dataset discovery services in two complementary ways: to suggest candidate datasets to authors during manuscript preparation or submission, and to surface related datasets to readers exploring a published work, supporting reproducibility and data reuse. For example, when a researcher searches for a publication, the system could identify potentially relevant datasets from the underlying \ac{SKG}, even when they are not directly linked to it, providing a discovery mechanism that complements keyword-based search. Such applications are particularly valuable because existing \acp{SKG} contain hundreds of millions of nodes yet only sparse publication–dataset links, as reflected in the extreme sparsity of MES and PubMed.

The choice of learning setting should be guided by the update frequency and growth rate of the target \ac{SKG}. The transductive setting, which yields the best performance, suits relatively stable resources for which the model can be periodically retrained offline, e.g., on a monthly or quarterly basis. The semi-inductive setting fits repositories that continuously receive new publications while recommending from an established dataset catalog. Finally, the fully inductive setting suits large aggregators such as \ac{OAG}, where both publications and datasets are added continuously and a complete retraining cycle at every update would be impractical; here, the ability to represent entirely unseen nodes, on both the publication and the dataset side, avoids the computational and operational cost of continuous retraining.

Lastly, deployment should be framed within the FAIR principles -- \textbf{F}indability, \textbf{A}ccessibility, \textbf{I}nteroperability, and \textbf{R}eusability~\cite{Wilkinson2016}. MM-SAN mainly contributes to Findability, by surfacing datasets that lack explicit links to publications, and thereby indirectly to Reusability. However, identifying a relevant dataset is only the first step towards its reuse: a dataset may require authentication or an access request, and its license may restrict redistribution, commercial use, or the creation of derivative resources. Recommendations should therefore be accompanied by up-to-date information on access conditions, licenses, and provenance, allowing researchers to distinguish immediately reusable resources from those requiring additional permissions. In future extensions, such metadata could also be incorporated as additional signals or constraints in the recommendation process. More generally, while the inductive capability reduces the need for frequent retraining, it does not remove the underlying data-management and governance requirements of real-world deployment.

\section{Final remarks}
\label{sec:remarks}

Data is a cornerstone of modern research, with significant efforts dedicated to its curation for learning and analytical tasks. 
However, data reuse remains limited due to the challenges of finding and assessing relevant datasets for specific research outputs. 
Dataset recommendation is essential to addressing this issue, yet it remains highly challenging due to the inherent noise, sparsity, and incompleteness of \acp{SKG}, which serve as the primary source for linking datasets and publications.  


We propose \ac{MM-SAN}, a multimodal graph representation learning method for enriching \acp{SKG} and recommending relevant datasets for publications. MM-SAN's effectiveness stems from a three-phase approach: (1) enhancing graph connectivity by introducing new nodes derived from textual attributes; (2) selecting relevant neighbors for a target node using random walks; and (3) combining neighbor representations for each type with cross attention to produce the final target representation. This architecture seamlessly integrates textual information with graph topology while not relying on complete metadata coverage.

We conduct extensive experiments across three settings (transductive, semi-inductive, fully inductive) and under ideal and real metadata conditions using two benchmarks extracted from the \ac{OAG}. Our evaluation emphasizes the often-overlooked importance of generating predictions for newly added items without requiring model retraining. The results demonstrate MM-SAN's effectiveness in all tested settings.

We analyzed the impact of missing textual metadata across all methods and datasets, focusing on \textbf{robustness}, \textbf{flexibility}, and \textbf{versatility}.
MM-SAN maintains strong performance across both large, sparse \acp{SKG} (i.e., PubMed) and smaller, well-curated datasets (i.e., MES), demonstrating robustness to noise and sparsity. 
Performance declines as dataset size increases, especially when textual descriptions are absent.
MM-SAN also excels in flexibility, performing well in transductive, semi-inductive, and fully inductive settings, making it suitable for dynamic real-world applications. 
MM-SAN emerges as the most effective model in most conditions, although text-only approaches remain stronger when high-quality textual metadata is fully available. Future work could explore mixture-of-experts architectures to better exploit such metadata. 
Additionally, we plan to extend MM-SAN to incorporate FAIR-related accessibility considerations into the recommendation process, investigating how access conditions, licensing restrictions, provenance, and interoperability-related metadata can be used to distinguish between relevant datasets and those that are also practically accessible and reusable. We also plan to extend this analysis to venue-based recommendation, examining how these FAIR-related characteristics may vary across different node types and connectivity patterns.


\section*{Acknowledgments}
This work is supported by the HEREDITARY Project, as part of the European Union’s Horizon Europe research and innovation programme under Grant Agreement No GA 101137074.

\clearpage
\acrodef{3G}[3G]{Third Generation Mobile System}
\acrodef{5S}[5S]{Streams, Structures, Spaces, Scenarios, Societies}
\acrodef{AAAI}[AAAI]{Association for the Advancement of Artificial Intelligence}
\acrodef{AAL}[AAL]{Annotation Abstraction Layer}
\acrodef{AAM}[AAM]{Automatic Annotation Manager}
\acrodef{ACLIA}[ACLIA]{Advanced Cross-Lingual Information Access}
\acrodef{ACM}[ACM]{Association for Computing Machinery}
\acrodef{ADSL}[ADSL]{Asymmetric Digital Subscriber Line}
\acrodef{ADUI}[ADUI]{ADministrator User Interface}
\acrodef{AIP}[AIP]{Archival Information Package}
\acrodef{AJAX}[AJAX]{Asynchronous JavaScript Technology and \acs{XML}}
\acrodef{ALU}[ALU]{Aritmetic-Logic Unit}
\acrodef{AMUSID}[AMUSID]{Adaptive MUSeological IDentity-service}
\acrodef{ANOVA}[ANOVA]{ANalysis Of VAriance}
\acrodef{ANSI}[ANSI]{American National Standards Institute}
\acrodef{AP}[AP]{Average Precision at Rank 1000}
\acrodef{APC}[APC]{AP Correlation}
\acrodef{API}[API]{Application Program Interface}
\acrodef{AR}[AR]{Address Register}
\acrodef{AS}[AS]{Annotation Service}
\acrodef{ASAP}[ASAP]{Adaptable Software Architecture Performance}
\acrodef{ASI}[ASI]{Annotation Service Integrator}
\acrodef{ASM}[ASM]{Annotation Storing Manager}
\acrodef{ASR}[ASR]{Automatic Speech Recognition}
\acrodef{ASUI}[ASUI]{ASsessor User Interface}
\acrodef{ATIM}[ATIM]{Annotation Textual Indexing Manager}
\acrodef{AUC}[AUC]{Area Under the ROC Curve}
\acrodef{AUI}[AUI]{Administrative User Interface}
\acrodef{AWARE}[AWARE]{Assessor-driven Weighted Averages for Retrieval Evaluation}
\acrodef{BANKS-I}[BANKS-I]{Browsing ANd Keyword Searching I}
\acrodef{BANKS-II}[BANKS-II]{Browsing ANd Keyword Searching II}
\acrodef{bpref}[bpref]{Binary Preference}
\acrodef{BNF}[BNF]{Backus and Naur Form}
\acrodef{BoE}[BoE]{Bag-of-Entities}
\acrodef{BoW}[BoW]{Bag-of-Words}
\acrodef{BoWE}[BoWE]{Bag-of-Word-Embeddings}
\acrodef{BRICKS}[BRICKS]{Building Resources for Integrated Cultural Knowledge Services}
\acrodef{CAN}[CAN]{Content Addressable Netword}
\acrodef{CAS}[CAS]{Content-And-Structure}
\acrodef{CBSD}[CBSD]{Component-Based Software Developlement}
\acrodef{CBSE}[CBSE]{Component-Based Software Engineering}
\acrodef{CB-SPE}[CB-SPE]{Component-Based \acs{SPE}}
\acrodef{CD}[CD]{Collaboration Diagram}
\acrodef{CD}[CD]{Compact Disk}
\acrodef{CENL}[CENL]{Conference of European National Librarians}
\acrodef{CIDOC CRM}[CIDOC CRM]{CIDOC Conceptual Reference Model}
\acrodef{CIR}[CIR]{Current Instruction Register}
\acrodef{CIRCO}[CIRCO]{Coordinated Information Retrieval Components Orchestration}
\acrodef{CG}[CG]{Cumulated Gain}
\acrodef{CLAIRE}[CLAIRE]{Combinatorial visuaL Analytics system for Information Retrieval Evaluation}
\acrodef{CLEF}[CLEF]{Conference and Labs of the Evaluation Forum}
\acrodef{CLIR}[CLIR]{Cross Language Information Retrieval}
\acrodef{CLSM}[CLSM]{Convolutional Latent Semantic Model}
\acrodef{CMS}[CMS]{Content Management System}
\acrodef{CMT}[CMT]{Campaign Management Tool}
\acrodef{CNN}[CNN]{Convolutional Neural Network}
\acrodef{CNR}[CNR]{Italian National Council of Research}
\acrodef{CO}[CO]{Content-Only}
\acrodef{COD}[COD]{Code On Demand}
\acrodef{CODATA}[CODATA]{Committee on Data for Science and Technology}
\acrodef{COLLATE}[COLLATE]{Collaboratory for Annotation Indexing and Retrieval of Digitized Historical Archive Material}
\acrodef{Conv-KNRM}[Conv-KNRM]{Convolutional Kernel-based Neural Ranking Model}
\acrodef{Co-PACRR}[Co-PACRR]{A Context-Aware Neural IR Model for Ad-hoc Retrieval}
\acrodef{CP}[CP]{Characteristic Pattern}
\acrodef{CPE}[CPE]{Control Processor Element}
\acrodef{CPU}[CPU]{Central Processing Unit}
\acrodef{CQL}[CQL]{Contextual Query Language}
\acrodef{CRP}[CRP]{Cumulated Relative Position}
\acrodef{CRUD}[CRUD]{Create--Read--Update--Delete}
\acrodef{CS}[CS]{Characteristic Structure}
\acrodef{CSM}[CSM]{Campaign Storing Manager}
\acrodef{CSS}[CSS]{Cascading Style Sheets}
\acrodef{CU}[CU]{Control Unit}
\acrodef{CUI}[CUI]{Client User Interface}
\acrodef{CV}[CV]{Cross-Validation}
\acrodef{DAFFODIL}[DAFFODIL]{Distributed Agents for User-Friendly Access of Digital Libraries}
\acrodef{DAO}[DAO]{Data Access Object}
\acrodef{DARE}[DARE]{Drawing Adequate REpresentations}
\acrodef{DARPA}[DARPA]{Defense Advanced Research Projects Agency}
\acrodef{DAS}[DAS]{Distributed Annotation System}
\acrodef{DB}[DB]{DataBase}
\acrodef{DBMS}[DBMS]{DataBase Management System}
\acrodef{DC}[DC]{Dublin Core}
\acrodef{DCI}[DCI]{Data Citation Index}
\acrodef{DCG}[DCG]{Discounted Cumulated Gain}
\acrodef{DCMI}[DCMI]{Dublin Core Metadata Initiative}
\acrodef{DCV}[DCV]{Document Cut--off Value}
\acrodef{DLI}[DLI]{Data-Literature Interlinking}
\acrodef{DD}[DD]{Deployment Diagram}
\acrodef{DDC}[DDC]{Dewey Decimal Classification}
\acrodef{DDS}[DDS]{Direct Data Structure}
\acrodef{DF}[DF]{Degrees of Freedom}
\acrodef{DFR}[DFR]{Divergence From Randomness}
\acrodef{DHT}[DHT]{Distributed Hash Table}
\acrodef{DI}[DI]{Digital Image}
\acrodef{DIKW}[DIKW]{Data, Information, Knowledge, Wisdom}
\acrodef{DIL}[DIL]{\acs{DIRECT} Integration Layer}
\acrodef{DiLAS}[DiLAS]{Digital Library Annotation Service}
\acrodef{DIRECT}[DIRECT]{Distributed Information Retrieval Evaluation Campaign Tool}
\acrodef{DKMS}[DKMS]{Data and Knowledge Management System}
\acrodef{DL}[DL]{Digital Library}
\acrodefplural{DL}[DL]{Digital Libraries}
\acrodef{DLMS}[DLMS]{Digital Library Management System}
\acrodef{DLOG}[DL]{Description Logics}
\acrodef{DLS}[DLS]{Digital Library System}
\acrodef{DLSS}[DLSS]{Digital Library Service System}
\acrodef{DO}[DO]{Digital Object}
\acrodef{DOI}[DOI]{Digital Object Identifier}
\acrodef{DOM}[DOM]{Document Object Model}
\acrodef{DoMDL}[DoMDL]{Document Model for Digital Libraries}
\acrodef{DPBF}[DPBF]{Dynamic Programming Best-First}
\acrodef{DR}[DR]{Data Register}
\acrodef{DRMM}[DRMM]{Deep Relevance Matching Model}
\acrodef{DRIVER}[DRIVER]{Digital Repository Infrastructure Vision for European Research}
\acrodef{DSKG}[DSKG]{Data Set Knowledge Graph}
\acrodef{DSSM}[DSSM]{Deep Structured Semantic Model}
\acrodef{DTD}[DTD]{Document Type Definition}
\acrodef{DVD}[DVD]{Digital Versatile Disk}
\acrodef{EAC-CPF}[EAC-CPF]{Encoded Archival Context for Corporate Bodies, Persons, and Families}
\acrodef{EAD}[EAD]{Encoded Archival Description}
\acrodef{EAN}[EAN]{International Article Number}
\acrodef{ECD}[ECD]{Enhanced Contenty Delivery}
\acrodef{ECDL}[ECDL]{European Conference on Research and Advanced Technology for Digital Libraries}
\acrodef{ECIR}[ECIR]{European Conference on Information Retrieval}
\acrodef{EDM}[EDM]{Europeana Data Model}
\acrodef{EG}[EG]{Execution Graph}
\acrodef{EHRs}[EHRs]{Electronic Health Records}
\acrodef{ELDA}[ELDA]{Evaluation and Language resources Distribution Agency}
\acrodef{ELRA}[ELRA]{European Language Resources Association}
\acrodef{EM}[EM]{Expectation Maximization}
\acrodef{EMMA}[EMMA]{Extensible MultiModal Annotation}
\acrodef{EPROM}[EPROM]{Erasable Programmable \acs{ROM}}
\acrodef{EQNM}[EQNM]{Extended Queueing Network Model}
\acrodef{ER}[ER]{Entity--Relationship}
\acrodef{ERR}[ERR]{Expected Reciprocal Rank}
\acrodef{ETL}[ETL]{Extract-Transform-Load}
\acrodef{FAST}[FAST]{Flexible Annotation Service Tool}
\acrodef{FFNN}[FFNN]{Feed Forward Neural Network}
\acrodef{FIFO}[FIFO]{First-In / First-Out}
\acrodef{FIRE}[FIRE]{Forum for Information Retrieval Evaluation}
\acrodef{FN}[FN]{False Negative}
\acrodef{FNR}[FNR]{False Negative Rate}
\acrodef{FOAF}[FOAF]{Friend of a Friend}
\acrodef{FORESEE}[FORESEE]{FOod REcommentation sErvER}
\acrodef{FP}[FP]{False Positive}
\acrodef{FPR}[FPR]{False Positive Rate}
\acrodef{GCN}[GCN]{Graph Convolutional Network}
\acrodef{GIF}[GIF]{Graphics Interchange Format}
\acrodef{GIR}[GIR]{Geografic Information Retrieval}
\acrodef{GAP}[GAP]{Graded Average Precision}
\acrodef{GLM}[GLM]{General Linear Model}
\acrodef{GLangModel}[GLM]{Generalized Language Model}
\acrodef{GLMM}[GLMM]{General Linear Mixed Model}
\acrodef{GMAP}[GMAP]{Geometric Mean Average Precision}
\acrodef{GNN}[GNN]{Graph Neural Network}
\acrodef{GoP}[GoP]{Grid of Points}
\acrodef{GPRS}[GPRS]{General Packet Radio Service}
\acrodef{gRBP}[gRBP]{Graded Rank-Biased Precision}
\acrodef{GRL}[GRL]{Graph Representation Learning}
\acrodef{GTIN}[GTIN]{Global Trade Item Number}
\acrodef{GUI}[GUI]{Graphical User Interface}
\acrodef{GW}[GW]{Gateway}
\acrodef{HCI}[HCI]{Human Computer Interaction}
\acrodef{HDS}[HDS]{Hybrid Data Structure}
\acrodef{HIR}[HIR]{Hypertext Information Retrieval}
\acrodef{HINT}[HiNT]{HIerarchical Neural maTching model}
\acrodef{HIT}[HIT]{Human Intelligent Task}
\acrodef{HITS}[HITS]{Hyperlink-Induced Topic Search}
\acrodef{HTML}[HTML]{HyperText Markup Language}
\acrodef{HTTP}[HTTP]{HyperText Transfer Protocol}
\acrodef{HSD}[HSD]{Honestly Significant Difference}
\acrodef{IAA}[IAA]{Inter Annotator Agreement}
\acrodef{ICA}[ICA]{International Council on Archives}
\acrodef{ICSU}[ICSU]{International Council for Science}
\acrodef{IDF}[IDF]{Inverse Document Frequency}
\acrodef{IDS}[IDS]{Inverse Data Structure}
\acrodef{IE}[IE]{Information Extraction}
\acrodef{IEEE}[IEEE]{Institute of Electrical and Electronics Engineers}
\acrodef{IEI}[IEI]{Istituto della Enciclopedia Italiana fondata da Giovanni Treccani}
\acrodef{IETF}[IETF]{Internet Engineering Task Force}
\acrodef{IMS}[IMS]{Information Management System}
\acrodef{IMSPD}[IMS]{Information Management Systems Research Group}
\acrodef{indAP}[indAP]{Induced Average Precision}
\acrodef{infAP}[infAP]{Inferred Average Precision}
\acrodef{INEX}[INEX]{INitiative for the Evaluation of \acs{XML} Retrieval}
\acrodef{INS-M}[INS-M]{Inverse Set Data Model}
\acrodef{INTR}[INTR]{Interrupt Register}
\acrodef{IP}[IP]{Internet Protocol}
\acrodef{IPSA}[IPSA]{Imaginum Patavinae Scientiae Archivum}
\acrodef{IR}[IR]{Information Retrieval}
\acrodef{IRON}[IRON]{Information Retrieval ON}
\acrodef{IRON2}[IRON$^2$]{Information Retrieval On aNNotations}
\acrodef{IRON-SAT}[IRON-SAT]{\acs{IRON} - Statistical Analysis Tool}
\acrodef{IRS}[IRS]{Information Retrieval System}
\acrodef{ISAD(G)}[ISAD(G)]{International Standard for Archival Description (General)}
\acrodef{ISBN}[ISBN]{International Standard Book Number}
\acrodef{ISIS}[ISIS]{Interactive SImilarity Search}
\acrodef{ISJ}[ISJ]{Interactive Searching and Judging}
\acrodef{ISO}[ISO]{International Organization for Standardization}
\acrodef{ITU}[ITU]{International Telecommunication Union }
\acrodef{ITU-T}[ITU-T]{Telecommunication Standardization Sector of \acs{ITU}}
\acrodef{IV}[IV]{Information Visualization}
\acrodef{JAN}[JAN]{Japanese Article Number}
\acrodef{JDBC}[JDBC]{Java DataBase Connectivity}
\acrodef{JMB}[JMB]{Java--Matlab Bridge}
\acrodef{JPEG}[JPEG]{Joint Photographic Experts Group}
\acrodef{JSON}[JSON]{JavaScript Object Notation}
\acrodef{JSP}[JSP]{Java Server Pages}
\acrodef{JTE}[JTE]{Java-Treceval Engine}
\acrodef{K-NRM}[K-NRM]{Kernel based Neural Ranking Model}
\acrodef{KDE}[KDE]{Kernel Density Estimation}
\acrodef{KG}[KG]{Knowledge Graph}
\acrodef{KLD}[KLD]{Kullback-Leibler Divergence}
\acrodef{KLAPER}[KLAPER]{Kernel LAnguage for PErformance and Reliability analysis}
\acrodef{LAM}[LAM]{Libraries, Archives, and Museums}
\acrodef{LAM2}[LAM]{Logistic Average Misclassification}
\acrodef{LAN}[LAN]{Local Area Network}
\acrodef{LD}[LD]{Linked Data}
\acrodef{LDA}[LDA]{Latent Dirichlet Allocation}
\acrodef{LEAF}[LEAF]{Linking and Exploring Authority Files}
\acrodef{LIDO}[LIDO]{Lightweight Information Describing Objects}
\acrodef{LIFO}[LIFO]{Last-In / First-Out}
\acrodef{LM}[LM]{Language Model}
\acrodef{LMT}[LMT]{Log Management Tool}
\acrodef{LOD}[LOD]{Linked Open Data}
\acrodef{LODE}[LODE]{Linking Open Descriptions of Events}
\acrodef{LpO}[LpO]{Leave-$p$-Out}
\acrodef{LRM}[LRM]{Local Relational Model}
\acrodef{LRU}[LRU]{Last Recently Used}
\acrodef{LS}[LS]{Lexical Signature}
\acrodef{LSM}[LSM]{Log Storing Manager}
\acrodef{LSTM-DSSM}[LSTM-DSSM]{LSTM Deep Structured Semantic Model}
\acrodef{LUG}[LUG]{Lexical Unit Generator}
\acrodef{MA}[MA]{Mobile Agent}
\acrodef{MA}[MA]{Moving Average}
\acrodef{MACS}[MACS]{Multilingual ACcess to Subjects}
\acrodef{MADCOW}[MADCOW]{Multimedia Annotation of Digital Content Over the Web}
\acrodef{MAD}[MAD]{Mean Assessed Documents}
\acrodef{MADP}[MADP]{Mean Assessed Documents Precision}
\acrodef{MADS}[MADS]{Metadata Authority Description Standard}
\acrodef{MAG}[MAG]{Microsoft Academic Graph}
\acrodef{MAKG}[MAKG]{Microsoft Academic Knowledge Graph}
\acrodef{MAP}[MAP]{Mean Average Precision at Rank 1000}
\acrodef{MARC}[MARC]{Machine Readable Cataloging}
\acrodef{MATTERS}[MATTERS]{MATlab Toolkit for Evaluation of information Retrieval Systems}
\acrodef{MDA}[MDA]{Model Driven Architecture}
\acrodef{MDD}[MDD]{Model-Driven Development}
\acrodef{MES}[MES]{European Marine Science}
\acrodef{METS}[METS]{Metadata Encoding and Transmission Standard}
\acrodef{MIDI}[MIDI]{Musical Instrument Digital Interface}
\acrodef{MIME}[MIME]{Multipurpose Internet Mail Extensions}
\acrodef{ML}[ML]{Machine Learning}
\acrodef{MLIA}[MLIA]{MultiLingual Information Access}
\acrodef{MM}[MM]{Machinery Model}
\acrodef{MMU}[MMU]{Memory Management Unit}
\acrodef{MODS}[MODS]{Metadata Object Description Schema}
\acrodef{MOF}[MOF]{Meta-Object Facility}
\acrodef{MP}[MP]{Markov Precision}
\acrodef{MPEG}[MPEG]{Motion Picture Experts Group}
\acrodef{MRD}[MRD]{Machine Readable Dictionary}
\acrodef{MRF}[MRF]{Markov Random Field}
\acrodef{MS}[MS]{Mean Squares}
\acrodef{MSAC}[MSAC]{Multilingual Subject Access to Catalogues}
\acrodef{MSE}[MSE]{Mean Square Error}
\acrodef{MT}[MT]{Machine Translation}
\acrodef{MV}[MV]{Majority Vote}
\acrodef{MVC}[MVC]{Model-View-Controller}
\acrodef{MM-SAN}[MM-SAN]{Multimodal Scholarly Attention Network}
\acrodef{NACSIS}[NACSIS]{NAtional Center for Science Information Systems}
\acrodef{NAP}[NAP]{Network processors Applications Profile}
\acrodef{NCP}[NCP]{Normalized Cumulative Precision}
\acrodef{nCG}[nCG]{Normalized Cumulated Gain}
\acrodef{nCRP}[nCRP]{Normalized Cumulated Relative Position}
\acrodef{nDCG}[nDCG]{Normalized Discounted Cumulated Gain}
\acrodef{nDCG@20}[nDCG@20]{Normalized Discounted Cumulated Gain at Cutoff 20}
\acrodef{nDCG@100}[nDCG@100]{Normalized Discounted Cumulated Gain at Cutoff 100}
\acrodef{NESTOR}[NESTOR]{NEsted SeTs for Object hieRarchies}
\acrodef{NER}[NER]{Named Entity Recognition}
\acrodef{NER+L}[NER+L]{Named Entity Recognition and Linking}
\acrodef{NeuIR}[NeuIR]{Neural Information Retrieval}
\acrodef{NEXI}[NEXI]{Narrowed Extended XPath I}
\acrodef{NII}[NII]{National Institute of Informatics}
\acrodef{NISO}[NISO]{National Information Standards Organization}
\acrodef{NIST}[NIST]{National Institute of Standards and Technology}
\acrodef{NLP}[NLP]{Natural Language Processing}
\acrodef{NP}[NP]{Network Processor}
\acrodef{NR}[NR]{Normalized Recall}
\acrodef{NS-M}[NS-M]{Nested Set Model}
\acrodef{NTCIR}[NTCIR]{NII Testbeds and Community for Information access Research}
\acrodef{NVSM}[NVSM]{Neural Vector Space Model}
\acrodef{OAG}[OAG]{OpenAIRE Graph}
\acrodef{OAI}[OAI]{Open Archives Initiative}
\acrodef{OAI-ORE}[OAI-ORE]{Open Archives Initiative Object Reuse and Exchange}
\acrodef{OAI-PMH}[OAI-PMH]{Open Archives Initiative Protocol for Metadata Harvesting}
\acrodef{OAIS}[OAIS]{Open Archival Information System}
\acrodef{OC}[OC]{Operation Code}
\acrodef{OCLC}[OCLC]{Online Computer Library Center}
\acrodef{OMG}[OMG]{Object Management Group}
\acrodef{OLAP}[OLAP]{On-Line Analytical Processing}
\acrodef{OO}[OO]{Object Oriented}
\acrodef{OODB}[OODB]{Object-Oriented \acs{DB}}
\acrodef{OODBMS}[OODBMS]{Object-Oriented \acs{DBMS}}
\acrodef{OPAC}[OPAC]{Online Public Access Catalog}
\acrodef{OS}[OS]{Open Science}
\acrodef{OQL}[OQL]{Object Query Language}
\acrodef{ORG}[OAG]{OpenAIRE Graph}
\acrodef{ORKG}[ORKG]{Open Research Knowledge Graph}
\acrodef{ORP}[ORP]{Open Relevance Project}
\acrodef{OSIRIS}[OSIRIS]{Open Service Infrastructure for Reliable and Integrated process Support}
\acrodef{P@1}[P@1]{Precision at Cut-off 1}
\acrodef{P@10}[P@10]{Precision at Cut-off 10}
\acrodef{P@20}[P@20]{Precision at Cut-off 20}
\acrodef{P2P}[P2P]{Peer-To-Peer}
\acrodef{PA}[PA]{Performance Analysis}
\acrodef{PAMT}[PAMT]{Pool-Assessment Management Tool}
\acrodef{PASM}[PASM]{Pool-Assessment Storing Manager}
\acrodef{PACRR}[PACRR]{A Position-Aware Neural IR Model for Relevance Matching}
\acrodef{PC}[PC]{Program Counter}
\acrodef{PCP}[PCP]{Pre-Commercial Procurement}
\acrodef{PCR}[PCR]{Peripherical Command Register}
\acrodef{PDA}[PDA]{Personal Digital Assistant}
\acrodef{PDF}[PDF]{Probability Density Function}
\acrodef{PDR}[PDR]{Peripherical Data Register}
\acrodef{PL2}[PL2]{Poisson estimation for randomness using Laplace succession for normalisation}
\acrodef{POI}[POI]{\acs{PURL}-based Object Identifier}
\acrodef{PoS}[PoS]{Part of Speech}
\acrodef{PPE}[PPE]{Programmable Processing Engine}
\acrodef{PREFORMA}[PREFORMA]{PREservation FORMAts for culture information/e-archives}
\acrodef{PRIMAmob-UML}[PRIMAmob-UML]{mobile \acs{PRIMA-UML}}
\acrodef{PRIMA-UML}[PRIMA-UML]{PeRformance IncreMental vAlidation in \acs{UML}}
\acrodef{PROM}[PROM]{Programmable \acs{ROM}}
\acrodef{PROMISE}[PROMISE]{Participative Research labOratory  for Multimedia and Multilingual Information Systems Evaluation}
\acrodef{pSQL}[pSQL]{propagate \acs{SQL}}
\acrodef{PUI}[PUI]{Participant User Interface}
\acrodef{PURL}[PURL]{Persistent \acs{URL}}
\acrodef{QA}[QA]{Question Answering}
\acrodef{QoS-UML}[QoS-UML]{\acs{UML} Profile for QoS and Fault Tolerance}
\acrodef{RAM}[RAM]{Random Access Memory}
\acrodef{RAMM}[RAM]{Random Access Machine}
\acrodef{RBO}[RBO]{Rank-Biased Overlap}
\acrodef{RBP}[RBP]{Rank-Biased Precision}
\acrodef{RDBMS}[RDBMS]{Relational \acs{DBMS}}
\acrodef{RDF}[RDF]{Resource Description Framework}
\acrodef{REST}[REST]{REpresentational State Transfer}
\acrodef{REV}[REV]{Remote Evaluation}
\acrodef{RFC}[RFC]{Request for Comments}
\acrodef{RIA}[RIA]{Reliable Information Access}
\acrodef{RMSE}[RMSE]{Root Mean Square Error}
\acrodef{RMT}[RMT]{Run Management Tool}
\acrodef{RNN}[RNN]{Recurrent Neural Network}
\acrodef{ROM}[ROM]{Read Only Memory}
\acrodef{ROMIP}[ROMIP]{Russian Information Retrieval Evaluation Seminar}
\acrodef{RoMP}[RoMP]{Rankings of Measure Pairs}
\acrodef{RoS}[RoS]{Rankings of Systems}
\acrodef{RP}[RP]{Relative Position}
\acrodef{RR}[RR]{Reciprocal Rank}
\acrodef{RSM}[RSM]{Run Storing Manager}
\acrodef{RST}[RST]{Rhetorical Structure Theory}
\acrodef{RT-UML}[RT-UML]{\acs{UML} Profile for Schedulability, Performance and Time}
\acrodef{SA}[SA]{Software Architecture}
\acrodef{SAL}[SAL]{Storing Abstraction Layer}
\acrodef{SAMT}[SAMT]{Statistical Analysis Management Tool}
\acrodef{SAN}[SAN]{Scholarly Attention Network}
\acrodef{SASM}[SASM]{Statistical Analysis Storing Manager}
\acrodef{SD}[SD]{Sequence Diagram}
\acrodef{SE}[SE]{Search Engine}
\acrodef{SEBD}[SEBD]{Convegno Nazionale su Sistemi Evoluti per Basi di Dati}
\acrodef{SFT}[SFT]{Satisfaction--Frustration--Total}
\acrodef{SIGIR}[SIGIR]{ACM SIGIR Conference on Research \& Development in Information Retrieval}
\acrodef{SIL}[SIL]{Service Integration Layer}
\acrodef{SIP}[SIP]{Submission Information Package}
\acrodef{SKG}[SKG]{Scholarly Knowledge Graph}
\acrodefplural{SKG}[SKGs]{Scholarly Knowledge Graphs}
\acrodef{SG}[SG]{Scholarly Graph}
\acrodef{SKOS}[SKOS]{Simple Knowledge Organization System}
\acrodef{SM}[SM]{Software Model}
\acrodef{SMART}[SMART]{System for the Mechanical Analysis and Retrieval of Text}
\acrodef{SoA}[SoA]{Service-oriented Architectures}
\acrodef{SOA}[SOA]{Strength of Association}
\acrodef{SOAP}[SOAP]{Simple Object Access Protocol}
\acrodef{SOM}[SOM]{Self-Organizing Map}
\acrodef{SOTA}[SOTA]{State-Of-The-Art}
\acrodef{SPE}[SPE]{Software Performance Engineering}
\acrodef{SPINA}[SPINA]{Superimposed Peer Infrastructure for iNformation Access}
\acrodef{SPLIT}[SPLIT]{Stemming Program for Language Independent Tasks}
\acrodef{SPOOL}[SPOOL]{Simultaneous Peripheral Operations On Line}
\acrodef{SQL}[SQL]{Structured Query Language}
\acrodef{SR}[SR]{Sliding Ratio}
\acrodef{SR}[SR]{Status Register}
\acrodef{SRNN}[SRNN]{Spatial Recurrent Neural Network}
\acrodef{SRU}[SRU]{Search/Retrieve via \acs{URL}}
\acrodef{SS}[SS]{Sum of Squares}
\acrodef{SSTF}[SSTF]{Shortest Seek Time First}
\acrodef{STAR}[STAR]{Steiner-Tree Approximation in Relationship graphs}
\acrodef{STON}[STON]{STemming ON}
\acrodef{TAC}[TAC]{Text Analysis Conference}
\acrodef{TBG}[TBG]{Time-Biased Gain}
\acrodef{TCP}[TCP]{Transmission Control Protocol}
\acrodef{TEL}[TEL]{The European Library}
\acrodef{TERRIER}[TERRIER]{TERabyte RetrIEveR}
\acrodef{TF}[TF]{Term Frequency}
\acrodef{TFR}[TFR]{True False Rate}
\acrodef{TN}[TN]{True Negative}
\acrodef{TO}[TO]{Transfer Object}
\acrodef{TP}[TP]{True Positve}
\acrodef{TPR}[TPR]{True Positive Rate}
\acrodef{TRAT}[TRAT]{Text Relevance Assessing Task}
\acrodef{TREC}[TREC]{Text REtrieval Conference}
\acrodef{TRECVID}[TRECVID]{TREC Video Retrieval Evaluation}
\acrodef{TTL}[TTL]{Time-To-Live}
\acrodef{UCD}[UCD]{Use Case Diagram}
\acrodef{UDC}[UDC]{Universal Decimal Classification}
\acrodef{uGAP}[uGAP]{User-oriented Graded Average Precision}
\acrodef{UI}[UI]{User Interface}
\acrodef{UML}[UML]{Unified Modeling Language}
\acrodef{UMT}[UMT]{User Management Tool}
\acrodef{UMTS}[UMTS]{Universal Mobile Telecommunication System}
\acrodef{UoM}[UoM]{Utility-oriented Measurement}
\acrodef{UPC}[UPC]{Universal Product Code}
\acrodef{URI}[URI]{Uniform Resource Identifier}
\acrodef{URL}[URL]{Uniform Resource Locator}
\acrodef{URN}[URN]{Uniform Resource Name}
\acrodef{USM}[USM]{User Storing Manager}
\acrodef{VA}[VA]{Visual Analytics}
\acrodef{VATE}[VATE$^2$]{Visual Analytics Tool for Experimental Evaluation}
\acrodef{VIRTUE}[VIRTUE]{Visual Information Retrieval Tool for Upfront Evaluation}
\acrodef{VD}[VD]{Virtual Document}
\acrodef{VIAF}[VIAF]{Virtual International Authority File}
\acrodef{VL}[VL]{Visual Language}
\acrodef{VoIP}[VoIP]{Voice over IP}
\acrodef{VS}[VS]{Visual Sentence}
\acrodef{W3C}[W3C]{World Wide Web Consortium}
\acrodef{WAN}[WAN]{Wide Area Network}
\acrodef{WHO}[WHO]{World Health Organization}
\acrodef{WLAN}[WLAN]{Wireless \acs{LAN}}
\acrodef{WP}[WP]{Work Package}
\acrodef{WS}[WS]{Web Services}
\acrodef{WSD}[WSD]{Word Sense Disambiguation}
\acrodef{WSDL}[WSDL]{Web Services Description Language}
\acrodef{WWW}[WWW]{World Wide Web}
\acrodef{XMI}[XMI]{\acs{XML} Metadata Interchange}
\acrodef{XML}[XML]{eXtensible Markup Language}
\acrodef{XPath}[XPath]{XML Path Language}
\acrodef{XSL}[XSL]{eXtensible Stylesheet Language}
\acrodef{XSL-FO}[XSL-FO]{\acs{XSL} Formatting Objects}
\acrodef{XSLT}[XSLT]{\acs{XSL} Transformations}
\acrodef{YAGO}[YAGO]{Yet Another Great Ontology}
\acrodef{YASS}[YASS]{Yet Another Suffix Stripper}
\bibliographystyle{ACM-Reference-Format}
\newpage
\bibliography{bibliography}
\section*{Appendix}

\renewcommand{\thesection}{A}
\renewcommand{\thesubsection}{\thesection.\arabic{subsection}}
\setcounter{subsection}{0}

\renewcommand{\thefigure}{A\arabic{figure}}
\setcounter{figure}{0}

\renewcommand{\thetable}{A\arabic{table}}
\setcounter{table}{0}
In this appendix, we provide additional details regarding some key aspects of our study. First, we report the hyperparameter configurations of all baseline methods: for each baseline, we list the complete set of hyperparameter values explored during grid search, while the best-performing configurations are reported in Section~\ref{sec:evaluation}.

Second, we present a component analysis of our pipeline. In particular, we investigate the contribution of two critical stages, the augmentation phase and the sampling phase, evaluating their individual impact on the overall performance, also when applied to the baseline methods.

Third, we report additional analyses on MM-SAN. We first evaluate MM-SAN with simpler encoders, replacing \texttt{node2vec} with \texttt{DeepWalk} and the transformer-based text encoders with mean-pooled GloVe embeddings, to assess how much the choice of encoders contributes to its effectiveness. We also analyze the sensitivity of MM-SAN to the number of neighbors sampled for each node type.

Finally, we evaluate MM-SAN on a more challenging variant of the task, in which authors rather than publications serve as query nodes, to assess whether it generalizes beyond publication-based dataset recommendation.

\subsection{Baselines hyperparameters}
We grouped the baseline methods into three categories based on their underlying architectural mechanism, rather than by data type alone, since this let us define a shared and meaningful hyperparameter search space for each group.

The first group consists of the variational graph autoencoders (VGAE and HVGAE), which learn probabilistic node representations and reconstruct the graph structure through an inner-product decoder.
The second group brings together the attention-based models (GAT, KGAT, HGT, HAN, and HGNN), all of which rely on a multi-head attention mechanism to weigh the contribution of neighboring nodes.
The third group covers non-attention message-passing models with explicit neighbor sampling (GraphSAGE, RGCN, and KGCN), which instead aggregate neighborhood information through fixed aggregation functions or relation-specific weight decompositions.

For each group, we report the hyperparameters explored via grid search, including learning rate, hidden/embedding dimension, number of layers, dropout, and weight decay, along with group-specific parameters such as the number of attention heads, the neighbor sampling size, and the aggregation strategy. The hyperparameter ranges explored for each baseline are summarized in Tables \ref{tab:vgae-hvgae}, \ref{tab:attention-based}, \ref{tab:message-passing}.
\begin{table*}[h]
\centering
\caption{Hyperparameters for variational graph autoencoder models. }
\label{tab:vgae-hvgae}
\resizebox{\textwidth}{!}{
\begin{tabular}{lcccccccc}
\toprule
\textbf{Model} & \textbf{Learning Rate} & \textbf{Embedding Dim. $L$} & \textbf{Layers}   & \textbf{Weight Decay ($\lambda$)} & \textbf{Optimizer} \\
\midrule
VGAE  & \{1e-5, 1e-4, 5e-4, 1e-3, 5e-3, 1e-2\} & \{16, 32, 64, 128\} & \{2, 3\}  & \{0, 1e-4, 1e-3\} & Adam \\
HVGAE & \{1e-6, 1e-5, \text{1e-4}, 5e-4, 1e-3\} & \{8, 16, 32, 64\} & \{2\} &  \{1e-4, \text{1e-3}, 1e-2\} & Adam \\
\bottomrule
\end{tabular}}
\end{table*}

\begin{table*}[h]
\centering
\caption{Hyperparameters for attention-based models. }
\label{tab:attention-based}
\resizebox{\textwidth}{!}{
\begin{tabular}{lccccccc}
\toprule
\textbf{Model} & \textbf{Learning Rate} & \textbf{Hidden Dim.} & \textbf{Layers} & \textbf{Heads} & \textbf{Neighbors} & \textbf{Dropout} & \textbf{Weight Decay}  \\
\midrule
GAT  & \{1e-5, 1e-4, 5e-4, 1e-3, 5e-3, 1e-2\} & \{64, 128, 256\} & \{2, 3\} & \{1, 2, 4, 8\} & -                          & \{0.0, 0.1, 0.3, 0.5\} & \{0, 1e-5, 1e-4\} \\
KGAT & \{1e-5, 1e-4, 5e-4, 1e-3, 5e-3, 1e-2\} & \{64, 128, 256\} & \{2, 3\} & \{1, 2, 4, 8\} & \{5, 10, 15\}              & \{0.0, 0.1, 0.3, 0.5\} & \{0, 1e-5, 1e-4\}  \\
HGT  & \{1e-5, 1e-4, 5e-4, 1e-3, 5e-3, 1e-2\} & \{64, 128, 256\} & \{2, 3\} & \{1, 2, 4, 8\} & \{5, 10, 15\}  & \{0.0, 0.1, 0.3, 0.5\} & \{0, 1e-5, 1e-4\} \\
HAN  & \{1e-5, 1e-4, 5e-4, 1e-3, 5e-3, 1e-2\} & \{64, 128, 256\} & \{2, 3\} & \{1, 2, 4, 8\} & -       & \{0.0, 0.1, 0.3, 0.5\} & \{0, 1e-5, 1e-4\} \\
HGNN & \{1e-5, 1e-4, 5e-4, 1e-3, 5e-3, 1e-2\} & \{64, 128, 256\} & \{2, 3\} & \{1, 2, 4, 8\} & \{5, 10, 15\}         & \{0.0, 0.1, 0.3, 0.5\} & \{0, 1e-5, 1e-4\}  \\
\bottomrule
\end{tabular}}
\end{table*}

\begin{table*}[h]
\centering
\caption{Hyperparameters for non-attention message-passing models.}
\label{tab:message-passing}
\resizebox{\textwidth}{!}{
\begin{tabular}{lccccccc}
\toprule
\textbf{Model} & \textbf{Learning Rate} & \textbf{Hidden Dim.} & \textbf{Layers} & \textbf{Neighbors} & \textbf{Aggregator / Num. Bases} & \textbf{Dropout} & \textbf{Weight Decay} \\
\midrule
GraphSAGE & \{1e-5, 1e-4, 5e-4, 1e-3, 5e-3, 1e-2\} & \{64, 128, 256\} & \{2, 3\} & \{5, 10, 15\} & Aggr.: \{mean, pool, LSTM\} & \{0.0, 0.1, 0.3, 0.5\} & \{0, 1e-5, 1e-4\} \\
RGCN      & \{1e-5, 1e-4, 5e-4, 1e-3, 5e-3, 1e-2\} & \{64, 128, 256\} & \{2, 3\} & \{5, 10, 15\} & Bases: \{0, 4, 8, 16\}       & \{0.0, 0.1, 0.3, 0.5\} & \{0, 1e-5, 1e-4\} \\
KGCN      & \{1e-5, 1e-4, 5e-4, 1e-3, 5e-3, 1e-2\} & \{64, 128, 256\} & \{2, 3\} & \{5, 10, 15\} & Aggr.: score-weighted& \{0.0, 0.1, 0.3, 0.5\} & \{0, 1e-5, 1e-4\} \\
\bottomrule
\end{tabular}}
\end{table*}

\subsection{Component analysis}
\label{sec:component}

We analyze the impact of the two main components of our pipeline: the augmentation and sampling stages. Specifically, we evaluate five baseline methods (GAT, SAGE, HAN, HGT, and HGNN) alongside MM-SAN on datasets with and without augmentation, to assess whether and to what extent augmentation affects their performance. We further investigate the impact of the sampling strategy by comparing the performance of the selected methods under the random-walk sampling configuration.
Since the Recall@10, nDCG@10, and MRR@10 metrics exhibit trends that closely mirror those at cutoff 5, the latter are considered indicative of the system’s overall performance and are therefore used as the primary basis for our analysis.

\begin{table}[t]
\footnotesize
\caption{Augmentation analysis. Recall@5 (R@5), nDCG@5 (N@5), and Mean Reciprocal Rank (MRR@5) scores on MES and PubMed under ideal conditions with and without augmentation. Best results are in \textbf{bold}, second-best are \underline{underlined}.}
\label{table:no_aug}
\resizebox{\textwidth}{!}{
\begin{tabular}{l|ccc|ccc||ccc|ccc}
\toprule
\multirow{3}{*}{Method} & \multicolumn{6}{c||}{\textbf{MES}} & \multicolumn{6}{c}{\textbf{PubMed}} \\\cline{2-13}
& \multicolumn{3}{c|}{Augmentation} & \multicolumn{3}{c||}{No Augmentation} 
& \multicolumn{3}{c|}{Augmentation} & \multicolumn{3}{c}{No Augmentation} \\\cline{2-13}
& R@5 & N@5 & MRR@5 & R@5 & N@5 & MRR@5 & R@5 & N@5 & MRR@5 & R@5 & N@5 & MRR@5 \\
\midrule
GAT    & 0.578 & 0.422 & \underline{0.443} & \underline{0.625} & \underline{0.475} & \textbf{0.462} & 0.228 & 0.165 & 0.198 & 0.241 & 0.181 & 0.203\\
SAGE   & 0.472 & 0.312 & 0.411 & 0.489 & 0.398 & 0.448 & 0.217 & 0.172 & 0.191 & 0.239 & 0.179 & \underline{0.204}\\
HGT    & 0.020 & 0.020 & 0.010 & 0.023 & 0.020 & 0.020 & 0.014 & 0.015 & 0.009 & 0.016 & 0.017 & 0.018\\
HAN    & 0.008 & 0.009 & 0.008 & 0.008 & 0.009 & 0.009 & 0.026 & 0.021 & 0.010 & 0.030 & 0.023 & 0.024\\
HGNN   & \underline{0.601} & \underline{0.438} & 0.406 & 0.581 & 0.438 & 0.378 & \underline{0.294} & \underline{0.222} & \underline{0.218} & \underline{0.284} & \textbf{0.202} & \textbf{0.209}\\
\midrule
MM-SAN & \textbf{0.727} & \textbf{0.546} & \textbf{0.484} & \textbf{0.642} & \textbf{0.476} & \underline{0.450} & \textbf{0.380} & \textbf{0.272} & \textbf{0.234} & \textbf{0.311} & \underline{0.201} & \text{0.181} \\
\bottomrule
\end{tabular}}
\end{table}


\paragraph{\textbf{Augmentation.}}
We first assess how the augmentation phase \circled{1} in Figure \ref{fig:san_architecture} -- specifically, the integration of entity and topic nodes to reduce overall sparsity -- affects system effectiveness. Table \ref{table:no_aug} presents the performance of six systems with (\texttt{augmentation} column) and without (\texttt{no augmentation} column) this phase on the MES and PubMed datasets. 
Moreover, we focus on the ideal transductive setup with full metadata availability, as it is the configuration in which all baselines perform their best, and consequently provides the clearest assessment of the augmentation impact.

For GAT and GraphSAGE, which were originally developed for homogeneous graphs, the introduction of multiple node types tends to degrade their effectiveness. These architectures implicitly assume neighborhoods composed of a single node type and a shared feature space across all nodes, so when applied to heterogeneous \acp{SKG} they struggle to properly model type-specific semantics and relations, leading to suboptimal representations and weaker recommendation performance. In other words, the added nodes increase connectivity but also introduce noise that these models are not designed to handle. 
MM-SAN and HGNN benefit from introducing new nodes, improving performance with the enriched graph. This suggests that the augmentation phase is particularly beneficial for systems that exploit topological features and are designed explicitly for heterogeneous graphs.
For HAN and HGT, removing the augmentation phase does not significantly affect their final performance.
In summary, the augmentation phase proves useful for models like MM-SAN and HGNN that are tailored for heterogeneous graphs and leverage topological features. However, it may not benefit -- and can even hinder -- models not explicitly designed to handle heterogeneous structures.

\paragraph{\textbf{Sampling.}}
\begin{table*}[t]
\caption{Sampling analysis. Comparison between original baselines and their variants enhanced with our sampling strategy. The table reports Recall@5 (R@5), nDCG@5 (N@5), and Mean Reciprocal Rank@5 (MRR@5) on the MES and PubMed datasets under transductive (Tran), semi-inductive (Semi), and inductive (Ind) settings. 100\% indicates full metadata availability. The Gain\% column shows the relative improvement achieved by our sampling method. Best results are in \textbf{bold}, second-best are \underline{underlined}.}
\footnotesize
\label{table:mes_pubmed_cutoff_5_gain_fixed}
\resizebox{\textwidth}{!}{
\begin{tabular}{llll|rrr|rrr|rrr|r}
\toprule
 Dataset & & Setting & Metric & GAT & GAT-RW & Gain \% & SAGE & SAGE-RW & Gain \% & HGNN & HGNN-RW & Gain \% & MM-SAN \\
\midrule
\multirow{9}{*}{MES [100\%]} & 
& \multirow{3}{*}{Tran}
& R@5   & 0.625 & \underline{0.700} & 12.00 & 0.489 & 0.610 & 24.74 & 0.581 & \underline{0.700} & 20.48 & \textbf{0.727} \\
&&& N@5   & 0.475 & \underline{0.520} & 9.47  & 0.398 & 0.510 & 28.14 & 0.438 & \underline{0.520} & 18.72 & \textbf{0.546} \\
&&& MRR@5 & 0.462 & \underline{0.473} & 2.38  & 0.448 & 0.470 & 4.91  & 0.378 & 0.390 & 3.17  & \textbf{0.484} \\
\cline{3-14}
&& \multirow{3}{*}{Semi}
& R@5   & 0.621 & \underline{0.700} & 12.72 & 0.446 & 0.600 & 34.53 & 0.587 & \underline{0.700} & 19.25 & \textbf{0.719} \\
&&& N@5   & 0.479 & 0.490 & 2.30  & 0.316 & 0.400 & 26.58 & 0.417 & \underline{0.510} & 22.30 & \textbf{0.524} \\
&&& MRR@5 & 0.454 & \underline{0.456} & 0.44  & 0.437 & 0.455 & 1.79  & 0.362 & 0.380 & 4.97  & \textbf{0.461} \\
\cline{3-14}
&& \multirow{3}{*}{Ind}
& R@5   & 0.590 & 0.622 & 5.42  & 0.439 & 0.512 & 16.63 & 0.582 & \underline{0.634} & 8.94  & \textbf{0.687} \\
&&& N@5   & 0.442 & \underline{0.513} & 16.06 & 0.315 & 0.346 & 9.84  & 0.422 & 0.475 & 12.56 & \textbf{0.519} \\
&&& MRR@5 & 0.434 & 0.437 & 0.69  & 0.417 & \underline{0.438} & 0.23  & 0.370 & 0.377 & 1.89  & \textbf{0.466} \\
\midrule\midrule
\multirow{9}{*}{PubMed [100\%]} & 
& \multirow{3}{*}{Tran}
& R@5   & 0.241 & 0.298 & 23.65 & 0.239 & 0.278 & 16.32 & 0.284 & \underline{0.301} & 5.96  & \textbf{0.380} \\
&&& N@5   & 0.181 & 0.202 & 11.60 & 0.179 & 0.201 & 12.29 & 0.202 & \underline{0.256} & 26.73 & \textbf{0.272} \\
&&& MRR@5 & 0.203 & 0.214 & 5.42  & 0.204 & 0.219 & 7.35  & 0.209 & \underline{0.225} & 7.66  & \textbf{0.234} \\
\cline{3-14}
&& \multirow{3}{*}{Semi}
& R@5   & 0.231 & 0.242 & 4.76  & 0.221 & 0.253 & 14.48 & 0.246 & \underline{0.271} & 10.16 & \textbf{0.373} \\
&&& N@5   & 0.171 & \underline{0.205} & 19.88 & 0.169 & 0.182 & 7.69  & 0.177 & 0.201 & 13.56 & \textbf{0.266} \\
&&& MRR@5 & 0.202 & 0.205 & 1.49  & 0.204 & 0.202 & -0.98 & \underline{0.222} & 0.213 & -4.05 & \textbf{0.232} \\
\cline{3-14}
&& \multirow{3}{*}{Ind}
& R@5   & 0.196 & 0.240 & 22.45 & 0.207 & 0.239 & 27.81 & 0.271 & \underline{0.320} & 18.08 & \textbf{0.344} \\
&&& N@5   & 0.134 & 0.186 & 38.81 & 0.133 & 0.201 & 51.13 & 0.199 & \textbf{0.267} & 34.17 & \underline{0.246} \\
&&& MRR@5 & 0.182 & 0.199 & 9.34  & 0.193 & \textbf{0.214} & 0.47  & 0.187 & 0.201 & 7.49  & \textbf{0.214} \\
\bottomrule
\end{tabular}}
\end{table*}
To test the effectiveness of our sampling phase (phase \circled{2} in Figure \ref{fig:san_architecture}), we reimplemented the best-performing baselines -- namely SAGE, GAT and HGNN -- and adapted them to incorporate our sampling strategy. 
In SAGE and GAT we replaced their random sampling strategy with our random walk-based one, while in HGNN we modified the random walk–based sampling to select walks and neighbors according to their semantic similarity to the target node.
The resulting performance scores for the MES and PubMed datasets at 100\% of available metadata are presented in Table \ref{table:mes_pubmed_cutoff_5_gain_fixed}. Similarly to the previous case, the sampling strategy does not rely on metadata availability; as a consequence, we provide the results for the ideal scenario with 100\% of available metadata. 
We find that integrating our sampling strategy into the baseline models leads to consistent performance gains when evaluated on real-world datasets like MES and PubMed. On MES, GraphSAGE benefits the most, achieving improvements of up to 35\% in Recall and 28\% in nDCG. On PubMed, GAT and GraphSAGE exhibit the largest boosts over their original implementations, with GraphSAGE achieving up to 51\% improvement in nDCG and GAT reaching 38\%. Despite these gains from the modified sampling strategy, MM-SAN remains the most effective model, consistently outperforming the adapted baselines in Recall, nDCG, and MRR. These results demonstrate that our sampling strategy yields consistent and meaningful improvements when applied to real-world \acp{SKG} with high incompleteness, node imbalance, sparseness, and dynamicity.

\subsection{Additional experiments}
\begin{table}[t]
\caption{Encoders analysis. The \textit{Simpler} column reports the performance of MM-SAN when the original encoders are replaced with simpler alternatives: DeepWalk for graph embeddings and mean-pooled GloVe embeddings for textual representations.}
\label{tab:simpler}
\scriptsize
\begin{tabular}{ll|rr||rr}
\toprule
& & \multicolumn{2}{c||}{\textbf{MES}} & \multicolumn{2}{c}{\textbf{PubMed}} \\
\cline{3-6}
Setting & Metric
& Simpler & MM-SAN
& Simpler & MM-SAN \\
\midrule
\multirow{3}{*}{Tran [100\%]}
& R@5    & 0.700 & \textbf{0.727} & 0.317 & \textbf{0.380} \\
& nDCG@5 & 0.520 & \textbf{0.546} & 0.219 & \textbf{0.272} \\
& MRR@5  & 0.453 & \textbf{0.484} & 0.217 & \textbf{0.234} \\
\midrule
\multirow{3}{*}{Ind [50\%]}
& R@5    & 0.449 & \textbf{0.526} & 0.179 & \textbf{0.242} \\
& nDCG@5 & 0.291 & \textbf{0.391} & 0.106 & \textbf{0.169} \\
& MRR@5  & 0.231 & \textbf{0.326} & 0.109 & \textbf{0.121} \\
\bottomrule
\end{tabular}
\end{table}

\subsubsection{Encoders analysis}\label{app:encoder}To assess the contribution of the encoding models employed in MM-SAN, we compared the original \texttt{node2vec} graph encoder and the Transformer-based text encoders (\texttt{Phrase-BERT} and \texttt{all-MiniLM-L6-v2}) against simpler alternatives. Specifically, we replaced \texttt{node2vec} with \texttt{DeepWalk} for graph representation learning and the Transformer-based text encoders with GloVe embeddings \cite{glove}, where textual representations were obtained by mean-pooling the embeddings of the words contained in each text.
The results are presented in Table \ref{tab:simpler}. The results show that the original encoders consistently outperform the simpler alternatives across both datasets and learning settings. In the transductive setting, MM-SAN improves R@5 from $0.700$ to $0.727$ on MES and from $0.317$ to $0.380$ on PubMed when using the original encoders. The difference becomes more pronounced in the inductive setting, where R@5 increases from $0.449$ to $0.526$ on MES and from $0.179$ to $0.242$ on PubMed. Similar improvements are observed for nDCG@5 and MRR@5. These results suggest that richer structural and contextual representations are particularly beneficial when only partial metadata is available. Nevertheless, the simpler encoders remain competitive, indicating that the overall MM-SAN framework remains effective even with less expressive representations.

\begin{figure}[t]
\centering
\includegraphics[width=0.8\textwidth]{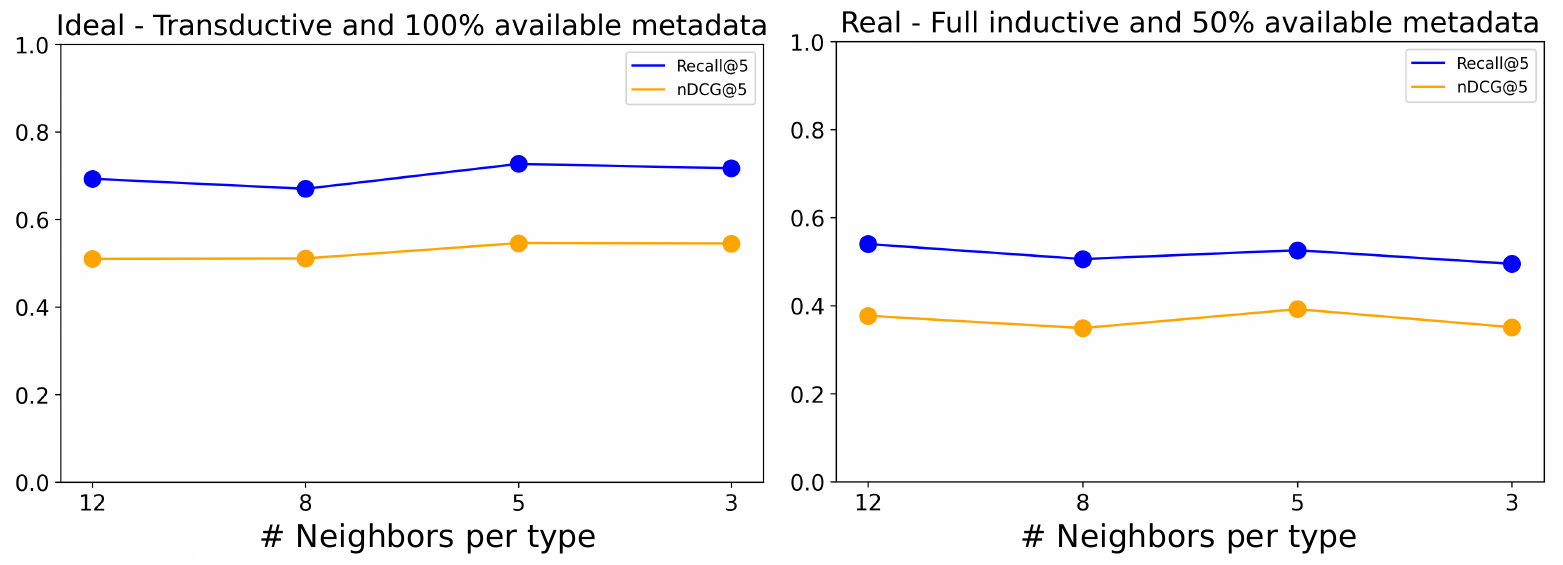}
\caption{Hyperparameter analysis: number of neighbors of a target node sampled for each node type.}\label{fig:neigh}
\Description{The SAN Architecture.}
\end{figure}
\subsubsection{Neighbors analysis}
We analyzed the impact of the number of sampled neighbors on MM-SAN's overall effectiveness. Figure~\ref{fig:neigh} reports the results on the MES dataset under two representative settings: an ideal scenario (transductive with 100\% metadata availability) and a more realistic scenario (fully inductive with 50\% metadata availability). In both cases, performance varies only slightly with the number of sampled neighbors and reaches its best overall performance when five neighbors per node type are selected. This configuration provides a good balance between representational richness and computational cost, supporting its use as the default sampling size. The limited sensitivity to the number of neighbors also suggests that MM-SAN's cross-attention mechanism can effectively identify relevant information within the sampled neighborhoods, contributing to its robustness across different graph conditions.

 \subsubsection{Dataset recommendation for authors}\label{app:authors}
\begin{table*}[t]
\centering
\caption{Dataset recommendation for authors results. Recall@$k$ (R@$k$), nDCG@$k$ (N@$k$) and Mean Reciprocal Rank@$k$ (MRR@$k$), with $k = 5$, on MES and PubMed in transductive (Tran), semi-inductive (Semi), and inductive (Ind) settings, and with \(\{100,50\}\% \) of datasets with available metadata. Best results are in \textbf{bold}, second-best are \underline{underlined}.}
\label{table:authors}
\resizebox{\textwidth}{!}{
\begin{tabular}{lll|rrrrrrr|rrrrrrr}
\toprule
 & & &
\multicolumn{7}{c|}{\textbf{MES $k=5$}} &
\multicolumn{7}{c}{\textbf{PubMed $k=5$}} \\
\text{Metadata \%} & \text{Setting} & \text{Metric} &
GAT & KGAT & SAGE & HGT & HAN & HGNN & MM-SAN &
GAT & KGAT & SAGE & HGT & HAN & HGNN & MM-SAN \\
\midrule
\multirow{9}{*}{100\%} & \multirow{3}{*}{Tran}
& R@$k$  & 0.120 & 0.098 & 0.105  & 0.010 & 0.008 & \underline{0.567} & \textbf{0.677}
& 0.081 & 0.018 &  0.098&  0.004& 0.003  & \underline{0.189} & \textbf{0.212}  \\
& & N@$k$  & 0.100 &  0.075& 0.095  & 0.004 &  0.005& \underline{0.422} & \textbf{0.539}
& 0.066 & 0.009 &  0.069& 0.001 & 0.001 & \underline{0.129}& \textbf{0.180} \\
& & MRR@$k$ & 0.118  & 0.078 & 0.091  & 0.006 & 0.005 &\underline{0.375} & \textbf{0.407}
&0.071 & 0.010 & 0.083 &  0.002&  0.001& \underline{0.101} & \textbf{0.106} \\
\cline{2-17}
 & \multirow{3}{*}{Semi}
& R@$k$  & 0.092  & 0.078 &  0.085& 0.008 & 0.007 & \underline{0.530} & \textbf{0.613}
& 0.078 & 0.010 & 0.082 &  0.001&0.001  &\underline{0.150}& \textbf{0.200} \\
& & N@$k$  & 0.076 & 0.055 & 0.053  &0.003 & 0.004 & \underline{0.382} & \textbf{0.446}
& 0.065 & 0.008 &  0.071& 0.000 &0.000  &\underline{0.100}& \textbf{0.123} \\
& & MRR@$k$ & 0.089  & 0.071 & 0.064 &0.004  &0.003  &\underline{0.332} & \textbf{0.391}
& 0.053&  0.008& 0.061 &  0.000&  0.001& \underline{0.109} & \textbf{0.113} \\\cline{2-17}
 & \multirow{3}{*}{Ind}
& R@$k$  &  0.065&  0.054&  0.064&  0.000& 0.000 & \underline{0.362} & \textbf{0.468}
& 0.068 & 0.010 &  0.062&  0.000&  0.000 &\underline{0.099}& \textbf{0.102} \\
& & N@$k$  & 0.041 & 0.033 & 0.042 &  0.000& 0.000 & \underline{0.276} & \textbf{0.363}
&0.041  &  0.006&  0.041& 0.000 & 0.000 &\underline{0.089}& \textbf{0.098} \\
& & MRR@$k$ &  0.054 &0.041  &0.051  & 0.000 & 0.000 &\underline{0.271} & \textbf{0.354}
& 0.041&  0.004&  0.056& 0.000 &  0.000& \underline{0.070} & \textbf{0.082} \\
\midrule\midrule
\multirow{9}{*}{50\%} & \multirow{3}{*}{Tran}
& R@$k$  & 0.101 &0.006 &0.024 & 0.006 &  0.005& \underline{0.410} & \textbf{0.456}
& 0.031 &  0.009& 0.043 &  0.003&  0.002& \underline{0.098}&  \textbf{0.106}\\
& & N@$k$  &  0.081 &0.004& 0.013  & 0.001 & 0.002 & \underline{0.234} & \textbf{0.272}
&0.010  &0.003  &  0.023& 0.000 & 0.002 &\underline{0.090}& \textbf{0.098} \\
& & MRR@$k$ & 0.094  & 0.004 &  0.019&  0.001& 0.001 &\underline{0.193} & \textbf{0.216}
&0.021 & 0.004 & 0.025 &  0.000& 0.000 &  \underline{0.091}& \textbf{0.103} \\
\cline{2-17}
 & \multirow{3}{*}{Semi}
& R@$k$  & 0.099 &0.005  & 0.013  & 0.002  &0.002  & \underline{0.328} & \textbf{0.356}
&  0.019&  0.004&0.020  &  0.001&  0.002& \underline{0.098}& \textbf{0.102} \\
& & N@$k$  &  0.072& 0.004 & 0.009 & 0.000  & 0.000  & \underline{0.190} & \textbf{0.212}
& 0.011  & 0.003 &  0.013&  0.000& 0.000 &\underline{0.064}& \textbf{0.092} \\
& & MRR@$k$ & 0.091  &  0.004& 0.010  & 0.001  & 0.000  &\underline{0.152} & \textbf{0.205}
& 0.012& 0.002 &0.015  & 0.000 & 0.000 & \underline{0.084} & \textbf{0.100} \\
\cline{2-17}
 & \multirow{3}{*}{Ind}
& R@$k$  & 0.095 &0.002  &0.003  &  0.001& 0.001 & \underline{0.208} & \textbf{0.307}
& 0.012 & 0.003  & 0.015 & 0.000 & 0.001 &\underline{0.077}&  \textbf{0.089}\\
& & N@$k$  & 0.054 & 0.001 & 0.002  &  0.000 & 0.000  & \underline{0.139} & \textbf{0.189}
& 0.008 & 0.002 &0.008  &0.000  & 0.000 &\underline{0.046}&  \textbf{0.073}\\
& & MRR@$k$ &  0.082 & 0.001 & 0.000  &  0.000 & 0.000  &\underline{0.106} & \textbf{0.130}
& 0.007 &  0.001& 0.009 & 0.000 & 0.000 & \underline{0.038} & \textbf{0.064} \\
\bottomrule
\end{tabular}}
\end{table*}

In this section, we evaluate the ability of MM-SAN to generalize to a more challenging dataset recommendation scenario in which authors, rather than publications, are used as query nodes. Compared to publications and datasets, author nodes do not directly benefit from the graph augmentation process, making the recommendation task inherently more difficult. Furthermore, author nodes are substantially sparser due to the absence of reliable author disambiguation, which often results in multiple nodes representing the same individual and, consequently, a large number of low-degree nodes. As a result, the information available from the query node itself is often limited, increasing the importance of effectively exploiting the surrounding graph structure.
In this context, neighborhood exploration becomes a key component of the recommendation process. MM-SAN mitigates the effects of author sparsity by leveraging the local graph structure whenever authors are connected to multiple research artifacts or when those artifacts are embedded in a richer context of related entities, such as organizations, topics, keywords, and other scholarly objects. Through its graph augmentation and neighborhood sampling mechanisms, MM-SAN can exploit indirect relational signals that would otherwise remain inaccessible from the author node alone, enabling more effective recommendations despite the sparsity of the author layer.

Following the same procedure adopted for the publication--dataset recommendation task, we constructed a second set of five edge splits for the author--dataset recommendation task, where authors serve as query nodes and datasets as candidate items to be ranked. A similar splitting strategy was applied to the complete set of author--dataset edges $\varepsilon_{d,a}$, resulting in training, validation, transductive test, semi-inductive test, and fully inductive test sets. For the MES dataset, the validation and each of the three test sets contain approximately 400 edges, while the training set contains approximately 2,800 edges. For PubMed, the validation and each test set contain approximately 1,000 edges, with approximately 20,000 edges assigned to training.

In Table~\ref{table:authors}, we report the results of the inductive models for dataset recommendation using authors as query nodes, considering settings with 100\% and 50\% of datasets with available metadata. Performance is evaluated using Recall@5, nDCG@5, and MRR@5. MM-SAN consistently achieves the best performance among the evaluated inductive methods. Approaches such as GAT, SAGE, HAN, HGT and KGAT perform substantially worse than MM-SAN, with only GAT and SAGE achieving Recall higher than 0.1 in the transductive setting on MES. This can be attributed to the high sparsity of author nodes and the resulting difficulty in identifying representative neighborhoods. Similarly to the publication-based task, HGNN is the strongest baseline both on MES and PubMed, suggesting that heterogeneous message passing is effective in exploiting the limited connectivity of the graph.

As the proportion of datasets with available metadata decreases, MM-SAN's performance also declines, as its learned representations become less informative.
When metadata is missing from a larger fraction of dataset neighbors, less semantic information can be propagated to the author representation, resulting in lower recommendation performance.
The inductive setting exhibits the largest performance degradation. On MES, MM-SAN's R@5 decreases from 0.677 to 0.468 with 100\% metadata, and from 0.456 to 0.307 with 50\% metadata. On PubMed, the corresponding decreases are from 0.212 to 0.102 and from 0.106 to 0.089, respectively. Similar trends are observed for nDCG@5 and MRR@5. 
The larger degradation observed in the inductive setting can be attributed to the combination of node novelty and author sparsity. For unseen authors, the model must infer relevant datasets from the available graph context surrounding the author, which is often sparse and only weakly informative. In this respect, when these nodes have few connections, and particularly when the available neighboring datasets provide limited metadata, the information available for constructing an informative representation is substantially reduced. This effect is especially pronounced in the fully inductive setting, where both the query authors and the target datasets are unseen during training.

Overall, these findings highlight the intrinsic difficulty of author-based dataset recommendation and the effectiveness of MM-SAN in making use of the limited structural and semantic information available in the \ac{SKG}.










\end{document}